\documentclass[10pt,twocolumn]{article}

\usepackage{amsmath}
\usepackage{amssymb}
\usepackage{graphics}

\newtheorem{axiom}{Axiom}

\title{\textbf{Quantum Theory From Five Reasonable Axioms}}

\author{Lucien Hardy\thanks{\texttt{hardy@qubit.org.  This is version
4}}\\
\textit{Centre for Quantum Computation,}\\
\textit{The Clarendon Laboratory,}\\
\textit{Parks road, Oxford OX1 3PU, UK}}

\begin{document}

\maketitle

\begin{abstract}
The usual formulation of quantum theory is based on rather obscure axioms
(employing complex Hilbert spaces, Hermitean operators, and the trace
formula for calculating probabilities).
In this paper it is shown that quantum theory can be derived from five
very reasonable axioms. The first four of these axioms are obviously
consistent with both quantum theory and classical probability theory.
Axiom 5 (which requires that there exist continuous reversible
transformations between pure states) rules out classical probability
theory.  If Axiom 5 (or even just the word ``continuous'' from Axiom 5)
is dropped then we obtain
classical probability theory instead.  This work provides some insight into
the reasons why quantum theory is the way it is.  For example, it explains
the need for complex numbers and where the trace formula comes from.
We also gain insight into the relationship between quantum theory and
classical probability theory.
\end{abstract}

\section{Introduction}

Quantum theory, in its usual formulation, is very abstract.  The basic
elements are vectors in a complex Hilbert space.  These determine
measured probabilities by means of the well known trace formula - a
formula which has no obvious origin.  It is natural to ask
why quantum theory is the way it is.  Quantum theory is simply a new
type of probability theory.  Like classical probability theory it can be
applied to a wide range of phenomena.  However, the rules of classical
probability theory can be determined by pure thought alone without any
particular appeal to experiment (though, of course, to develop
classical probability theory, we do employ some basic intuitions about
the nature of the world).  Is the same true of quantum theory?  Put
another way, could a 19th century theorist have developed quantum theory
without access to the empirical data that later became available to his
20th century descendants? In this paper it will be shown that quantum theory
follows from
five very reasonable axioms which might well have been posited without
any particular access to empirical data.  We will not recover any
specific form of
the Hamiltonian from the axioms since that belongs to particular
applications of quantum theory (for example - a set of interacting spins
or the motion of a particle in one dimension). Rather we will recover
the basic structure of quantum theory along with the most general type
of quantum evolution possible.  In addition we will only deal with the
case where there are a finite or countably infinite number of
distinguishable states corresponding to a finite or countably infinite
dimensional Hilbert space. We will not deal with continuous dimensional
Hilbert spaces.

The basic setting we will consider is one in which we have preparation
devices, transformation devices, and measurement devices. Associated
with each preparation will be a state defined in the following way:
\begin{description}
\item[\bf The state] associated with a particular preparation is
defined to be (that thing represented by) any mathematical object that
can be used to determine the
probability associated with the outcomes of any measurement that may be
performed on a system prepared by the given preparation.
\end{description}
Hence, a list of all probabilities pertaining to all possible measurements that
could be made would certainly represent the state.  However, this would
most likely over determine the state.  Since most physical theories have
some structure, a smaller set of probabilities
pertaining to a set of carefully chosen measurements may be
sufficient to determine the state.  This is the case in
classical probability theory and quantum theory.
Central to the axioms are two integers $K$ and $N$ which
characterize the type of system being considered.
\begin{itemize}
\item The {\it number of degrees of freedom}, $K$, is defined as the
minimum number of probability measurements needed to determine the
state, or, more
roughly, as the number of real parameters required to specify the state.
\item The {\it dimension}, $N$, is defined as the
maximum number of states that can be reliably distinguished from one
another in a single shot measurement.
\end{itemize}
We will only consider the case where the number of distinguishable
states is finite or countably infinite.
As will be shown below, classical probability theory has $K=N$ and quantum
probability theory has $K=N^2$ (note we do not assume that states
are normalized).

The five
axioms for quantum theory (to be stated again, in context, later) are
\begin{description}
\item[Axiom 1] {\it Probabilities}.  Relative frequencies (measured by
taking the proportion of times a particular outcome is observed)
tend to the same value (which we call the probability) for any case
where a given measurement is performed on a ensemble of $n$ systems
prepared by some given preparation in the limit as $n$ becomes infinite.
\item[Axiom 2] {\it Simplicity}. $K$ is determined by a function of
$N$ (i.e. $K=K(N)$) where $N=1,2,\dots$ and where, for each
given $N$, $K$ takes the minimum value consistent with the axioms.
\item[Axiom 3] {\it Subspaces}. A system whose state is constrained to
belong to an $M$
dimensional subspace (i.e. have support on only $M$ of a set of $N$ possible
distinguishable states) behaves like a system of dimension $M$.
\item[Axiom 4]  {\it Composite systems}. A composite system consisting of
subsystems $A$ and $B$ satisfies $N=N_AN_B$ and $K=K_AK_B$
\item[Axiom 5] {\it Continuity}. There exists a continuous reversible
transformation on a system between any two pure states of that
system.
\end{description}
The first four axioms are consistent with classical probability theory
but the fifth is not (unless the word ``continuous'' is dropped).
If the last axiom is dropped then, because of the
simplicity axiom, we obtain
classical probability theory (with $K=N$) instead of quantum theory
(with $K=N^2$).
It is very striking that we have here a set of axioms for quantum theory
which have the property that if a
single word is removed -- namely the word ``continuous'' in Axiom 5 --
then we obtain classical probability theory instead.

The basic idea of the proof is simple.  First we show how the state can
be described by a real vector, ${\bf p}$, whose entries are
probabilities and that the
probability associated with an arbitrary measurement is given by a linear
function, ${\bf r}\cdot {\bf p}$, of this vector (the vector ${\bf r}$
is associated with the measurement).  Then we show that we must have $K=N^r$
where $r$ is
a positive integer and that it follows from the simplicity axiom that
$r=2$ (the $r=1$ case being ruled out by Axiom 5). We consider the
$N=2$, $K=4$ case and recover quantum
theory for a two dimensional Hilbert space. The subspace axiom is then
used to construct quantum theory for general $N$.  We also obtain the
most general evolution of the state consistent with the axioms and show
that the state of a composite system can be represented by a positive
operator on the tensor product of the Hilbert spaces of the subsystems.
Finally, we show obtain the rules for updating the state after a
measurement.

This paper is organized in the following way.  First we will describe
the type of situation we wish to consider (in which we have preparation
devices, state transforming devices, and measurement devices).  Then we
will describe classical probability theory and quantum
theory.  In particular it will be shown how quantum theory can be put in
a form similar to classical probability theory.  After that we will
forget both classical and quantum probability theory and show how they
can be obtained from the axioms.

Various authors have set up axiomatic formulations of quantum
theory, for example see references
\cite{birkoff,mackey,piron,ludwig,mielnik,lande,fivel,accardi,landsman,cmw}
(see
also \cite{gleason,kochen,pitowsky}). Much of this work is in the quantum logic
tradition.  The advantage of the present work is that there
are a small number of simple axioms, these axioms can easily
be motivated without any particular appeal to experiment,
and the mathematical methods required to obtain quantum theory from
these axioms are very straightforward (essentially just linear algebra).

\section{Setting the Scene}

We will begin by describing the type of experimental situation we wish
to consider (see Fig.~1).  An experimentalist has three types of device.
One is a preparation device.  We can think of it as preparing physical
systems in some state.
\begin{figure*}[t]
{\includegraphics{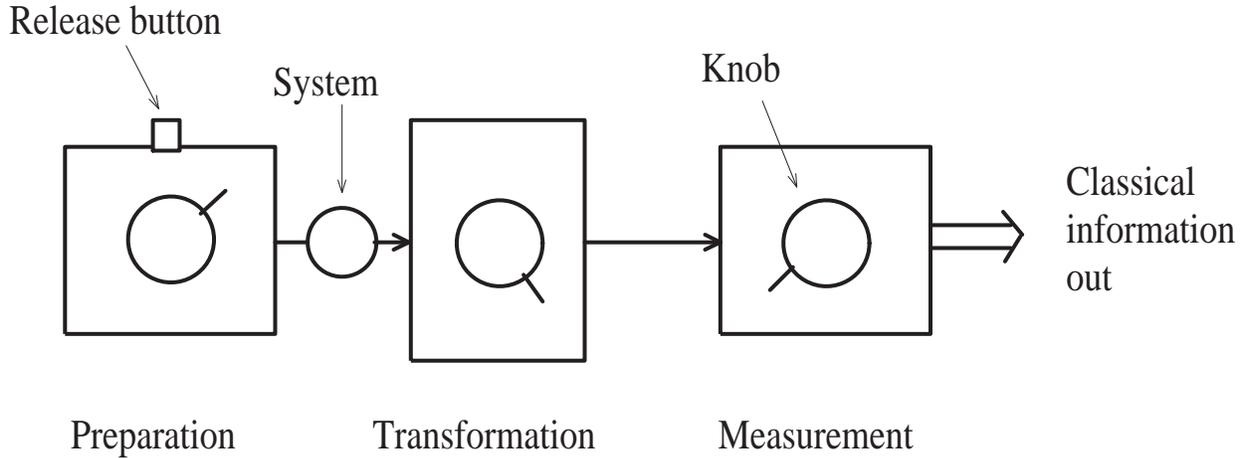}}
\caption{The situation considered consists of a preparation device with
a knob for varying the state of the system produced and a release button
for releasing the system, a transformation device for transforming the
state (and a knob to vary this transformation), and a measuring
apparatus for measuring the state (with a knob to vary what is measured)
which outputs a classical number.}
\end{figure*}
It has on it a number of knobs which can be
varied to change the state prepared.  The system is released by
pressing a button. The system passes through the second device.
This device can transform the
state of the system.  This device has knobs on it which can be
adjusted to effect different transformations (we might think of these as
controlling fields which effect the system). We can allow the system
to pass through a number of devices of this type.  Unless otherwise
stated, we will assume the transformation devices are set to allow the
system through unchanged.
Finally, we have a
measurement apparatus. This also has knobs on it which can be adjusted
to determine what measurement is being made.  This device outputs a
classical number.  If no system is incident on the device (i.e.
because the button on the preparation device was not pressed)
then it outputs a 0 (corresponding to a null outcome). If there is
actually a physical system incident (i.e when the release button is
pressed and the transforming device has not absorbed the system)
then the device outputs a number $l$
where $l=1$ to $L$ (we will call these  non-null outcomes).
The number of possible classical
outputs, $L$, may depend on what is being measured (the settings of the
knobs).

The fact that we allow null events means that we will not impose the
constraint that states are normalized.  This turns out to be a useful
convention.  It may appear that requiring the existence of null events
is an additional assumption.  However, it follows from the subspace
axiom that we can arrange to have a null outcome.
We can associate the non-null outcomes with a certain subspace and the
null outcome with the complement subspace.  Then we can restrict
ourselves to preparing only mixtures of states which are in the non-null
subspace (when the button is pressed) with states which are in the
null subspace (when the button is not pressed).

The situation described here is quite generic.  Although we have described
the set up as if the system were moving along one dimension, in fact the
system
could equally well be regarded as remaining stationary whilst being
subjected to transformations and measurements. Furthermore, the system
need not be localized but could be in several locations. The
transformations could be due to controlling fields or simply due to the
natural evolution of the system. Any physical experiment, quantum,
classical or other, can be viewed as an experiment of the type described here.

\section{Probability measurements}

We will consider only measurements of probability since all other
measurements (such as expectation values) can be calculated from
measurements of probability.  When, in this paper, we refer to a {\it
measurement} or a {\it probability measurement} we mean, specifically, a
measurement of the probability that the outcome belongs to some subset
of the non-null outcomes with a given setting of the knob on the
measurement apparatus.  For example, we could measure the probability
that the outcome is $l=1$ or $l=2$ with some given setting.

To perform a measurement we need a large number of identically prepared
systems.

A measurement returns a single real number (the probability)
between 0 and 1.  It is possible to perform many measurements at once.
For example, we could simultaneously measure [the probability the
outcome is $l=1$] and [the probability the outcome is $l=1$ or $l=2$]
with a given knob setting.

\section{Classical Probability Theory}

A classical system will have available to it a number, $N$, of distinguishable
states.  For example, we could consider a ball that can be in one of $N$
boxes.  We will call these distinguishable states the basis states.
Associated with each basis state will be the probability, $p_n$, of finding the
system in that state if we make a measurement. We can write
\begin{equation}
  {\bf p} = \left(
  \begin{matrix} p_1 \\ p_2 \\ p_3 \\ \vdots \\ p_N \end{matrix} \right)
\end{equation}
This vector can be regarded as describing the state of the system. It
can be determined by measuring $N$ probabilities and so $K=N$.
Note that we do not assume that the state is normalized (otherwise we
would have $K=N-1$).

The state ${\bf p}$ will belong to a convex set $S$.  Since the set is
convex it will have a subset of extremal states. These are the states
\begin{equation}\label{cpure}
{\bf p}_1= \left(
\begin{matrix} 1 \\ 0 \\ 0 \\ \vdots \\ 0 \end{matrix} \right) \qquad
{\bf p}_2= \left(
\begin{matrix} 0 \\ 1 \\ 0 \\ \vdots \\ 0 \end{matrix} \right) \qquad
{\bf p}_3= \left(
\begin{matrix} 0 \\ 0 \\ 1 \\ \vdots \\ 0 \end{matrix} \right) \qquad
{\rm etc.}
\end{equation}
and the state
\begin{equation}
{\bf p}_{\rm null}= {\bf 0} =
\left( \begin{matrix} 0 \\ 0 \\ 0 \\ \vdots \\ 0 \end{matrix} \right)
\end{equation}
The state ${\bf 0}$ is the null state (when the system is not present).
We define the set of pure states to consist of all extremal states
except the null state.  Hence, the states in (\ref{cpure}) are the pure
states.  They correspond to the system definitely being in one of the
$N$ distinguishable states.  A general state can be written as a convex
sum of the pure states and the null state and this gives us the exact
form of the set $S$.  This is always a polytope (a shape having flat
surfaces and a finite number of vertices).

We will now consider measurements.
Consider a measurement of the probability that the system is in the
basis state $n$.  Associated with this probability measurement is the
vector ${\bf r}_n$ having a 1 in position $n$ and 0's elsewhere. At
least for these cases the measured probability is given by
\begin{equation}\label{cpmeas}
p_{\rm meas} = {\bf r}\cdot{\bf p}
\end{equation}
However, we can consider more general types of probability measurement and
this formula will still hold. There are two ways in which we can
construct more general types of measurement:
\begin{enumerate}
\item We can perform a measurement in which we decide with probability
$\lambda$ to measure ${\bf r}_A$ and with probability $1-\lambda$ to
measure ${\bf r_B}$.  Then we will obtain a new measurement vector ${\bf
r}=\lambda {\bf r}_A +(1-\lambda) {\bf r}_B $.
\item We can add the results of two compatible probability measurements
and therefore add the corresponding measurement vectors.
\end{enumerate}
An example of the second is the probability measurement that the state
is basis state 1 or basis state 2 is given by the measurement vector
${\bf r}_1+{\bf r}_2$. From linearity, it is clear that the formula
(\ref{cpmeas}) holds for such more general measurements.

There must exist a measurement in which we simply check to see that the
system is present (i.e. not in the null state).  We denote this by ${\bf
r}^I$.  Clearly
\begin{equation}
{\bf r}^I=\sum_n {\bf r}_n =\left( \begin{matrix} 1 \\ 1 \\ 1 \\ \vdots
\\ 1 \end{matrix} \right)
\end{equation}
Hence $0\leq {\bf r}^I.{\bf p} \leq 1$ with normalized states saturating
the upper bound.

With a given setting of the knob on the measurement device there will be
a certain number of distinct non-null outcomes labeled $l=1$ to $L$.
Associated with each outcome will be a measurement vector ${\bf r}_l$.
Since, for normalized states, one non-null outcome must happen we have
\begin{equation}\label{sumls}
\sum_{l=1}^L {\bf r}_l = {\bf r}^I
\end{equation}
This equation imposes a constraint on any measurement vector. Let
allowed measurement vectors ${\bf r}$ belong to the set $R$. This
set is
clearly convex (by virtue of 1. above).  To fully determine $R$ first
consider the set $R^+$ consisting of all vectors which can be written
as a sum of the basis measurement vectors, ${\bf r}_n$,
each multiplied by a positive number. For
such vectors ${\bf r}\cdot{\bf p}$ is necessarily greater than 0 but may
also be greater than 1.  Thus, elements of $R^+$ may be too long to
belong to $R$.  We need a way of picking out those
elements of $R^+$ that also belong to $R$. If we can perform the
probability measurement ${\bf r}$ then, by (\ref{sumls})
we can also perform the probability measurement
$\overline{\bf r}\equiv {\bf r}^I - {\bf r}$. Hence,
\begin{equation}
{\rm Iff} \quad {\bf r},\overline{\bf r}\in R^+ \quad {\rm and} \quad
{\bf r}+\overline{\bf r}={\bf r}^I \quad{\rm then}\quad {\bf r},\overline{\bf r}\in R
\end{equation}
This works since it implies that ${\bf r}\cdot{\bf p} \leq 1$ for all ${\bf
p}$ so that ${\bf r}$ is not too long.

Note that the Axioms 1 to 4 are satisfied but Axiom 5 is not since there
are a finite number of pure states.  It is easy to show that reversible
transformations take pure states to pure states (see Section
\ref{axiomfive}). Hence a continuous reversible transformation will take
a pure state along a continuous path through the pure states which is
impossible here since there are only a finite number of pure states.

\section{Quantum Theory}\label{qtheory}

Quantum theory can be summarized by the following rules
\begin{description}
\item[States] The state is represented by a positive (and therefore Hermitean)
operator $\hat{\rho}$ satisfying $0\leq {\rm tr}(\hat{\rho})\leq 1$.
\item[Measurements] Probability measurements are represented by a
positive operator $\hat{A}$.  If $\hat{A}_l$ corresponds to outcome $l$
where $l=1$ to $L$ then
\begin{equation}
\sum_{l=1}^L \hat{A}_l = \hat{I}
\end{equation}
\item[Probability formula] The probability obtained when the measurement
$\hat{A}$ is made on the state $\hat{\rho}$ is
\begin{equation}\label{traceeqn}
p_{\rm meas}={\rm tr}(\hat{A}\hat{\rho})
\end{equation}
\item[Evolution] The most general evolution is given by the
superoperator $\$ $
\begin{equation}
\hat{\rho}\rightarrow \$ (\rho)
\end{equation}
where $\$ $
\begin{itemize}
\item Does not increase the trace.
\item Is linear.
\item Is completely positive.
\end{itemize}
\end{description}
This way of presenting quantum theory is rather condensed.  The
following notes should provide some clarifications
\begin{enumerate}
\item It is, again, more convenient not to impose normalization. This,
in any case, more accurately models what happens in real experiments when
the quantum system is often missing for some portion of the ensemble.
\item The most general type of measurement in quantum theory is a POVM
(positive operator valued measure).  The operator $\hat{A}$ is an
element of such a measure.
\item Two classes of superoperator are of particular interest.  If $\$ $
is reversible (i.e. the inverse $\$^{-1} $ both exists and belongs to
the allowed set of transformations)
then it will take pure states to pure states and
corresponds to unitary evolution.  The von Neumann projection
postulate takes the state $\hat{\rho}$ to the state
$\hat{P}\hat{\rho}\hat{P}$ when the outcome corresponds to the
projection operator $\hat{P}$.  This is a special case of a
superoperator evolution in which the trace of $\hat{\rho}$
decreases.
\item It has been shown by Krauss \cite{krauss} that one need only
impose the three
listed constraints on $\$ $ to fully constrain the possible types of quantum
evolution.  This includes unitary evolution and von Neumann projection
as already stated, and it also includes the evolution of an open system
(interacting with an environment). It is sometimes stated that the
superoperator should preserve the trace.  However, this is an
unnecessary constraint which makes it impossible to use the superoperator
formalism to describe von Neumann projection \cite{Nielsenchuang}.
\item The constraint that $\$ $ is completely positive imposes  not only
that $\$ $ preserves the positivity of $\hat{\rho}$ but also
that $ \$_A\otimes\hat{I}_B$ acting on any element of a tensor product
space also preserves positivity for any dimension of $B$.
\end{enumerate}

This is the usual formulation.  However, quantum theory can be recast in
a form more similar to classical probability theory. To do this we note
first that the space of Hermitean operators which act on a $N$
dimensional complex Hilbert space can be spanned by $N^2$ linearly
independent projection
operators $\hat{P}_k$ for $k=1$ to $K=N^2$.  This is clear since a general
Hermitean operator can be represented as a matrix. This matrix has $N$
real numbers along the diagonal and ${1\over2}N(N-1)$ complex numbers
above the diagonal making a total of $N^2$ real numbers.  An example of
$N^2$ such projection operators will be given later.  Define
\begin{equation}
\hat{\bf P}=\left(
\begin{matrix} \hat{P}_1 \\ \hat{P}_2 \\ \vdots \\ \hat{P}_K
\end{matrix} \right)
\end{equation}
Any Hermitean matrix can be written as a sum of these projection
operators times real numbers, i.e. in the form ${\bf a}\cdot\hat{\bf P}$
where $\bf a$ is a real vector ($\bf a$ is unique since the operators
$\hat{P}_k$ are linearly independent).
Now define
\begin{equation}
{\bf p}_S = {\rm tr}(\hat{\bf P} \hat{\rho})
\end{equation}
Here the subscript $S$ denotes \lq state\rq.  The $k$th component of
this vector is equal to the probability obtained when $\hat{P}_k$ is
measured on $\hat{\rho}$.
The vector ${\bf p}_S$ contains the same information as the state
$\hat{\rho}$
and can therefore be regarded as an alternative way of representing the
state.  Note that $K=N^2$ since it takes $N^2$ probability measurements
to determine
${\bf p}_S$ or, equivalently, $\hat{\rho}$. We define ${\bf r}_M$
through 
\begin{equation}\label{ArP}
\hat{A} = {\bf r}_M\cdot\hat{\bf P}
\end{equation}
The subscript $M$ denotes \lq measurement\rq. The vector ${\bf r}_M$ is
another way of representing the measurement $\hat{A}$.  If we
substitute (\ref{ArP}) into the trace formula (\ref{traceeqn}) we obtain
\begin{equation}\label{Qrp}
p_{\rm meas}= {\bf r}_M\cdot {\bf p}_S
\end{equation}
We can also define
\begin{equation}
{\bf p}_M = {\rm tr}(\hat{A}\hat{\bf P})
\end{equation}
and ${\bf r}_S$ by
\begin{equation}\label{rhoPr}
\hat{\rho}=\hat{\bf P}\cdot {\bf r}_S
\end{equation}
Using the trace formula (\ref{traceeqn}) we obtain
\begin{equation}\label{Qpr}
p_{\rm meas}= {\bf p}_M\cdot{\bf r}_S={\bf r}^T_M D {\bf r}_S
\end{equation}
where $T$ denotes transpose and $D$ is the $K\times K$ matrix with
real elements given by
\begin{equation}\label{QDij}
D_{ij}={\rm tr}(\hat{P}_i\hat{P}_j)
\end{equation}
or we can write $D={\rm tr}(\hat{\bf P}\hat{\bf P}^T)$.
From (\ref{Qrp},\ref{Qpr}) we obtain
\begin{equation}\label{QpDr}
{\bf p}_S=D{\bf r}_S
\end{equation}
and
\begin{equation}
{\bf p}_M=D^T {\bf r}_M
\end{equation}
We also note that
\begin{equation}
D=D^T
\end{equation}
though this would not be the case  had we chosen different spanning sets
of projection operators for the state operators and measurement operators.
The inverse $D^{-1}$ must exist (since the projection operators are
linearly independent).  Hence, we can also write
\begin{equation}
p_{\rm meas}= {\bf p}_M^T D^{-1} {\bf p}_S
\end{equation}

The state can be represented by an ${\bf r}$-type vector or a
${\bf p}$-type vector
as can the measurement.  Hence the subscripts $M$ and $S$ were
introduced.  We will sometimes drop these subscripts when it is clear
from the context whether the vector is a state or measurement vector.
We will stick to the convention of having measurement vectors on the
left and state vectors on the right as in the above formulae.

We define ${\bf r}^I$ by
\begin{equation}
\hat{I}={\bf r}^I\cdot\hat{\bf P}
\end{equation}
This measurement gives the probability of a non-null event. Clearly we
must have $0\leq {\bf r}^I\cdot{\bf p} \leq 1$ with normalized states
saturating the upper bound.
We can also define the measurement which tells us whether the state is
in a given subspace.  Let $\hat{I}_W$ be the projector into an $M$
dimensional subspace $W$.  Then the corresponding ${\bf r}$ vector is
defined by $\hat{I}_W={\bf r}^{I_W}\cdot\hat{\bf P}$.  We will
say that a state $\bf p$ is in the subspace $W$ if
\begin{equation}
{\bf r}^{I_W}\cdot{\bf p}={\bf r}^I\cdot{\bf p}
\end{equation}
so it only has support in $W$.  A system in which the state is always
constrained to an $M$-dimensional subspace will behave as an $M$
dimensional system in accordance with Axiom 3.

The transformation $\hat{\rho}\rightarrow \$ (\hat{\rho})$
of $\hat{\rho}$ corresponds to the following transformation for
the state vector ${\bf p}$:
\begin{eqnarray*}
{\bf p}
& = & {\rm tr}(\hat{\bf P}\hat{\rho})            \\
&\rightarrow & {\rm tr} (\hat{\bf P} \$(\hat{\rho})) \\
& = & {\rm tr} (\hat{\bf P} \$(\hat{\bf P}^T D^{-1} {\bf p})) \\
& = & Z {\bf p}
\end{eqnarray*}
where equations (\ref{rhoPr},\ref{QpDr}) were used in the third line and
$Z$ is a $K\times K$ real matrix given by
\begin{equation}\label{transprho}
Z={\rm tr}(\hat{\bf P}\$(\hat{\bf P})^T) D^{-1}
\end{equation}
(we have used the linearity property of $\$ $). Hence, we see that a linear
transformation in $\hat{\rho}$ corresponds to a linear transformation in ${\bf
p}$. We will say that $Z\in\Gamma$.

Quantum theory can now be summarized by the following rules
\begin{description}
\item[States] The state is given by a real vector ${\bf p}\in S$ with
$N^2$ components.
\item[Measurements] A measurement is represented by a real vector ${\bf
r}\in R$ with $N^2$ components.
\item[Probability measurements] The measured probability if measurement
${\bf r}$ is performed on state ${\bf p}$ is
\[ p_{\rm meas}= {\bf r}\cdot {\bf p}  \]
\item[Evolution] The evolution of the state is given by ${\bf
p}\rightarrow Z {\bf p}$ where $Z\in\Gamma$ is a real matrix.
\end{description}
The exact nature of the sets $S$, $R$ and $\Gamma$ can be deduced from
the equations relating these real vectors and matrices to their
counterparts in the usual quantum formulation.  We will show that these
sets can also be deduced from the
axioms. It has been noticed by various other authors that the state can
be represented by the probabilities used to determine it
\cite{wootters,stefan}.

There are various ways of choosing a set of $N^2$ linearly independent
projections operators $\hat{ P}_k$ which span the space of Hermitean
operators.  Perhaps the simplest way is the following. Consider an $N$
dimensional complex Hilbert space with an orthonormal basis set
$|n\rangle$ for $n=1$ to $N$.  We can define $N$ projectors
\begin{equation}\label{nbasis}
     |n\rangle\langle n|
\end{equation}
Each of these belong to one-dimensional subspaces formed from the
orthonormal basis set.  Define
\[ |mn\rangle_x={1\over\sqrt{2}}(|m\rangle + |n\rangle) \]
\[ |mn\rangle_y={1\over\sqrt{2}}(|m\rangle + i|n\rangle) \]
for $m < n$.  Each of these vectors has support on a two-dimensional
subspace formed from the orthonormal basis set.  There are ${1\over 2}
N(N-1)$ such two-dimensional subspaces.  Hence we can define
$N(N-1)$ further projection operators
\begin{equation}\label{mpmn}
 |mn\rangle_x\langle mn| ~~{\rm and}~~ |mn\rangle_y\langle mn|
\end{equation}
This makes a total of $N^2$ projectors. It is clear that these
projectors are linearly independent.

Each projector corresponds to one degree of freedom.  There is one
degree of freedom associated with each one-dimensional subspace $n$, and a
further two degrees of freedom associated with each two-dimensional
subspace $mn$.  It is possible, though not actually the case in quantum
theory, that there are further
degrees of freedom associated with each three-dimensional subspace and
so on.  Indeed, in general, we can write
\begin{equation}
\begin{array}{rcl}
K &= & Nx_1 + {1\over 2!}N(N-1)x_2 \\[3pt]
{} & {} &+ {1\over3!}N(N-1)(N-2)x_3 + \dots
\end{array}
\end{equation}
We will call the vector ${\bf x}=(x_1,x_2,\dots)$ the {\it signature} of
a particular probability theory.  Classical probability theory has
signature ${\bf x}_{\rm Classical}=(1,0,0,\dots)$ and quantum theory has
signature ${\bf x}_{\rm Quantum}=(1,2,0,0,\dots)$.  We will show that
these signatures are respectively picked out by Axioms 1 to 4 and Axioms
1 to 5.  The signatures ${\bf x}_{\rm Reals}=(1,1,0,0,\dots)$ of real
Hilbert space quantum theory and ${\bf x}_{\rm
Quaternions}=(1,4,0,0,\dots)$ of quaternionic quantum theory are ruled out.

If we have a composite system consisting of subsystem $A$ spanned by
$\hat{P}^A_i$ ($i=1$ to $K_A$) and $B$ spanned by
$\hat{P}^B_j$ ($j=1$ to $K_B$) then
$\hat{P}_i^A\otimes\hat{P}^B_j$ are linearly independent and span the
composite system.  Hence, for the composite system we have $K=K_AK_B$.
We also have $N=N_AN_B$.  Therefore Axiom 4 is satisfied.

The set $S$ is convex. It contains the null state $\bf 0$ (if the system
is never present) which is an extremal state.  Pure states are
defined as extremal states other than the null state (since they are
extremal they cannot be written as a convex sum of other states as we
expect of pure states).  We know that a  pure state can be represented
by a normalized vector $|\psi\rangle$.  This is specified by $2N-2$ real
parameters ($N$ complex numbers minus overall phase and minus
normalization).  On the other hand, the full set of normalized states is
specified by $N^2-1$ real numbers.  The surface of the set of normalized
states must therefore be $N^2-2$ dimensional.  This means that, in
general, the pure states are of lower dimension than the the surface of
the convex set of normalized states.  The only exception to this is the
case $N=2$ when the surface of the convex set is 2-dimensional and the
pure states are specified by two real parameters.  This case is
illustrated by the Bloch sphere. Points on the surface of the Bloch
sphere correspond to pure states.

In fact the $N=2$ case will play a particularly important role later so
we will now develop it a little further.  There will be four projection
operators spanning the space of Hermitean operators which we can choose
to be
\begin{equation}
\hat{P}_1=|1\rangle\langle 1|
\end{equation}
\begin{equation}
\hat{P}_2=|2\rangle\langle 2|
\end{equation}
\begin{equation}
\hat{P}_3=(\alpha|1\rangle+\beta|2\rangle)
(\alpha^*\langle 1|+\beta^*\langle 2|)
\end{equation}
\begin{equation}
\hat{P}_4=(\gamma|1\rangle+\delta|2\rangle)
(\gamma^*\langle 1|+\delta^*\langle 2|)
\end{equation}
where $|\alpha|^2+|\beta|^2=1$ and $|\gamma|^2+|\delta|^2=1$.
We have chosen the second pair of projections to be more general than
those defined in (\ref{mpmn}) above since we will need to consider this
more general case later.
We can calculate $D$ using (\ref{QDij})
\begin{equation}
D=\left( \begin{matrix}
       1 & 0 & 1-|\beta|^2 & 1-|\delta|^2 \\
       0 & 1 &  |\beta|^2 &   |\delta|^2  \\
     1-|\beta|^2 & |\beta|^2 & 1 & |\alpha\gamma^* +\beta\delta^*|^2 \\
     1-|\delta|^2 & |\delta|^2 & |\alpha\gamma^* +\beta\delta^*|^2 & 1
   \end{matrix} \right)
\end{equation}
We can write this as
\begin{equation}\label{QtwodD}
D=\left( \begin{array}{cccc} 1 &  0 &  1-a & 1-b \\
                             0 &  1 &   a  &  b  \\
                             1-a&  a &  1 &  c  \\
                            1-b &  b &  c &  1   \end{array} \right)
\end{equation}
where $a$ and $b$ are real with $\beta=\sqrt{a} \exp(i\phi_3)$,
$\delta=\sqrt{b} \exp(\phi_4)$, and
$c=|\alpha\gamma^* +\beta\delta^*|^2$. We can choose $\alpha$ and
$\gamma$ to be real (since the phase is included in the definition of
$\beta$ and $\delta$).  It then follows that
\begin{equation}\label{phase}
\begin{array}{ll}
\!\!c= & \!\!\!\!\!1-a-b+2ab \\[1pt]
\!\!                & \!\!\!\!\!
+2 \cos(\phi_4-\phi_3) \sqrt{ab(1-a)(1-b)}
\end{array}
\end{equation}
Hence, by varying the complex phase associated with $\alpha$, $\beta$, $\gamma$
and $\delta$ we find that
\begin{equation}\label{cpccm}
c_-< c < c_+
\end{equation}
where
\begin{equation}\label{Qroots}
c_{\pm}\equiv 1-a-b+2ab \pm 2\sqrt{ab(1-a)(1-b)}
\end{equation}
This constraint is equivalent to the condition ${\rm Det}(D) > 0$.
Now, if we are given a particular $D$ matrix of the form (\ref{QtwodD})
then we can go backwards to the usual quantum formalism though we must
make some arbitrary choices for the phases.
First we use (\ref{phase}) to calculate $\cos(\phi_4-\phi_3)$.  We can
assume that $0\leq \phi_4-\phi_3 \leq \pi$ (this corresponds to
assigning $i$ to one of the roots $\sqrt{-1}$).  Then we can assume
that $\phi_3=0$.  This fixes $\phi_4$.  An example of this second choice
is when we assign the state ${1\over\sqrt{2}}(|+\rangle+|-\rangle)$
(this has real coefficients) to spin along the $x$ direction for a spin
half particle.  This is arbitrary since we have rotational symmetry about
the $z$ axis.  Having calculated $\phi_3$ and $\phi_4$ from the elements
of $D$ we can now calculate $\alpha$, $\beta$, $\gamma$, and $\delta$
and hence we can obtain $\hat{\bf P}$. We can then
calculate $\hat{\rho}$,
$\hat{A}$ and $ \$ $ from ${\bf p}$, ${\bf r}$, and $Z$ and use the
trace formula.  The arbitrary choices for phases do not change any
empirical predictions.

\section{Basic Ideas and the Axioms}

We will now forget quantum theory and classical probability theory and
rederive them from the axioms.  In this section we will introduce the
basic ideas and the axioms in context.

\subsection{Probabilities}

As mentioned earlier, we will consider only measurements of probability
since all other measurements can be reduced to probability measurements.
We first need to ensure that it makes sense to talk of probabilities. To
have a probability we need two things.  First we need a way of preparing
systems (in Fig. 1 this is accomplished by the first two boxes) and
second, we need a way of measuring the systems (the third box in Fig.
1).  Then, we measure the number of cases, $n_+$, a particular outcome is
observed when a given measurement is performed on an ensemble of $n$
systems each prepared by a given preparation.  We define
\begin{equation}
{\rm prob}_+ = \lim_{n\rightarrow {\rm \infty}} {n_+\over n}
\end{equation}
In order for any theory of probabilities to make sense ${\rm prob}_+$ must take the
same value for any such infinite ensemble of systems prepared by a given
preparation.  Hence, we assume

\begin{axiom} {\rm Probabilities.}  Relative frequencies
(measured by taking the proportion of times a particular outcome is observed)
tend to the same value (which we call the probability) for any case
where a given measurement is performed on an ensemble of $n$ systems
prepared by some given preparation in the limit as $n$ becomes infinite.
\end{axiom}

With this axiom we can begin to build a probability theory.

Some additional comments are appropriate here.  There are various
different interpretations of probability: as frequencies, as
propensities, the Bayesian approach, etc.  As stated, Axiom 1 favours
the frequency approach.  However, it it equally possible to
cast this axiom in keeping with the other approaches \cite{schack}.
In this paper we are principally interested in deriving the structure of
quantum theory rather than solving the interpretational problems with
probability theory and so we will not try to be sophisticated with regard to
this matter.  Nevertheless, these are important questions which deserve
further attention.

\subsection{The state}

We can introduce the notion that the system is described by a state.
Each preparation will have a state associated with it. We
define the state to be (that thing represented by)
any mathematical object which can be used to
determine the probability for any measurement that could possibly be
performed on the system when prepared by the associated preparation. It
is possible to associate a state with a preparation because Axiom 1
states that these probabilities depend on the preparation and not on the
particular ensemble being used.  It follows from this definition of a
state that one way of representing the state is by a list of all
probabilities for all measurements that could possibly
be performed. However, this
would almost certainly be an over complete specification of the state
since most
physical theories have some structure which relates different measured
quantities. We expect that we will be able to consider a subset
of all possible measurements to determine the state.
Hence, to determine the state we need to make a number
of different measurements on
different ensembles of identically prepared systems.
A certain minimum number of appropriately chosen measurements will be
both necessary and sufficient to determine the state. Let this number be
$K$.  Thus, for each setting, $k=1$ to $K$, we will measure a
probability $p_k$ with an appropriate setting of the knob on the
measurement apparatus.
These $K$ probabilities can be represented by a column vector
${\bf p}$ where
\begin{equation}
  {\bf p} = \left(
  \begin{matrix} p_1 \\ p_2 \\ p_3 \\ \vdots \\ p_K \end{matrix} \right)
\end{equation}
Now, this vector contains just sufficient information to determine the state
and the state must contain just sufficient information to determine this
vector (otherwise it could not be used to predict probabilities for
measurements).  In other words, the state and this vector are
interchangeable and hence we can use ${\bf p}$ as a way of representing
the state of the system.
We will call $K$ the number of degrees of freedom associated with the
physical system.  We will not assume that the physical system is always
present.  Hence, one of the $K$ degrees of freedom can be associated
with normalization and therefore $K\geq 1$.

\subsection{Fiducial measurements}

We will call the probability measurements
labeled by $k=1$ to $K$ used in determining the state the {\it fiducial}
measurements.  There is no reason to suppose that this set is unique.  It
is possible that some other fiducial set could also be used to determine
the state.  

\subsection{Measured probabilities}

Any probability that can be measured (not just the
fiducial ones) will be determined by some function $f$ of
the state ${\bf p}$.  Hence,
\begin{equation}
p_{\rm meas}= f({\bf p})
\end{equation}
For different measurements the function will, of course, be different.
By definition, measured probabilities are between 0 and 1.
\[    0\leq p_{\rm meas} \leq 1 \]
This must be true since probabilities are measured by taking the proportion
of cases in which a particular event happens in an ensemble.

\subsection{Mixtures}

Assume that
the preparation device is in the hands of Alice.  She can decide randomly
to prepare a state ${\bf p}_A$ with probability $\lambda$ or a state
${\bf p}_B$ with probability $1-\lambda$.  Assume that she records this choice
but does not tell the person, Bob say, performing the measurement.
Let the state corresponding to this preparation be ${\bf p_C}$.  Then
the probability Bob measures will be the convex
combination of the two cases, namely
\begin{equation}\label{convexf}
f({\bf p}_C) = \lambda f({\bf p}_A)
+(1-\lambda) f({\bf p}_B)
\end{equation}
This is clear since Alice could subsequently reveal which
state she had prepared for each event in the ensemble providing two
sub-ensembles.  Bob could then check his data was consistent for each
subensemble.  By Axiom 1, the probability measured for each subensemble
must be the same as that which would have been measured for any
similarly prepared ensemble and hence (\ref{convexf}) follows.

\subsection{Linearity}

Equation (\ref{convexf}) can be applied to the fiducial measurements
themselves.  This gives
\begin{equation}\label{convexp}
{\bf p}_C = \lambda {\bf p}_A+(1-\lambda){\bf p}_B
\end{equation}
This is clearly true since it is true by (\ref{convexf}) for each component.

Equations (\ref{convexf},\ref{convexp}) give
\begin{equation}\label{convexpf}
f(\lambda {\bf p}_A+(1-\lambda){\bf p}_B)
= \lambda f({\bf p}_A)+(1-\lambda) f({\bf p}_B)
\end{equation}
This strongly suggests that the function $f(\cdot)$ is linear.
This is indeed the case and a proof is given in Appendix 1.
Hence, we can write
\begin{equation}\label{formrp}
p_{\rm meas} = {\bf r} \cdot {\bf p}
\end{equation}
The vector ${\bf r}$ is associated with the measurement.  The $k$th fiducial
measurement is the measurement which picks out the $k$th component of
$\bf p$.  Hence, the fiducial measurement vectors are
\begin{equation}\label{fidM}
{\bf r}^1= \left(
\begin{matrix} 1 \\ 0 \\ 0 \\ \vdots \\ 0 \end{matrix} \right) \quad
{\bf r}^2= \left(
\begin{matrix} 0 \\ 1 \\ 0 \\ \vdots \\ 0 \end{matrix} \right) \quad
{\bf r}^3= \left(
\begin{matrix} 0 \\ 0 \\ 1 \\ \vdots \\ 0 \end{matrix} \right) \quad
{\rm etc.}
\end{equation}

\subsection{Transformations}

We have discussed the role of the preparation device and the measurement
apparatus.  Now we will discuss the state transforming device (the
middle box in Fig.\ 1). If some system with state ${\bf p}$ is incident on this
device its state will be transformed to some new state ${\bf g}({\bf p})$.  It
follows from Eqn (\ref{convexf}) that this transformation must be
linear.  This is clear since we can apply the proof in the Appendix 1 to
each component of ${\bf g}$.  Hence, we can write the effect of the
transformation device as
\begin{equation}
{\bf p} \rightarrow Z {\bf p}
\end{equation}
where $Z$ is a $K\times K$ real matrix describing the effect of the
transformation.

\subsection{Allowed states, measurements, and transformations}

We now have states represented by $\bf p$, measurements represented by
$\bf r$, and transformations represented by $Z$.
These will each belong to some set
of physically allowed states, measurements and transformations. Let these
sets of allowed elements be $S$, $R$ and $\Gamma$.  Thus,
\begin{equation}
{\bf p} \in S
\end{equation}
\begin{equation}
{\bf r} \in R
\end{equation}
\begin{equation}
Z \in \Gamma
\end{equation}
We will use the axioms to determine the nature of these sets. It
turns out (for relatively obvious reasons) that each of these sets is convex.

\subsection{Special states}\label{specialstates}

If the release button on Fig.\ 1 is never pressed then all the fiducial
measurements will yield 0.  Hence, the null state ${\bf p}_{\rm null}={\bf 0}$
can be prepared and therefore ${\bf 0}\in S$.

It follows from (\ref{convexp}) that the set $S$ is convex. It is also
bounded since the entries of ${\bf p}$ are bounded by 0 and 1. Hence,
$S$ will have an extremal set $S_{\rm extremal}$ (these are the vectors in $S$
which cannot be written as a convex sum of other vectors in $S$).
We have  $\bf 0\in S_{\rm extremal}$ since the entries in the vectors
$\bf p$ cannot be negative.  We define the set of pure states $S_{\rm
pure}$ to be the set of all extremal states except $\bf 0$.
Pure states are clearly special in some way. They represent states which
cannot be interpreted as a mixture.  A driving intuition in this work
is the idea that pure states represent definite states of the system.

\subsection{The identity measurement}

The probability of a non-null outcome is given by summing up all the
non-null outcomes with a given setting of the knob on the measurement
apparatus (see Fig 1).  The non-null outcomes are labeled by $l=1$ to
$L$.
\begin{equation}\label{nonnull}
p_{\rm non-null} = \sum_{l=1}^L {\bf r}_l \cdot {\bf p}
={\bf r}^I \cdot {\bf p}
\end{equation}
where ${\bf r}_l$ is the measurement vector corresponding to outcome
$l$ and
\begin{equation}\label{identM}
{\bf r}^I = \sum_{l=1}^L {\bf r}_l
\end{equation}
is called the identity measurement.

\subsection{Normalized and unnormalized states}

If the release button is never pressed we prepare the state $\bf 0$.
If the release button is always pressed (i.e for every event in the
ensemble) then we will say ${\bf p}\in S_{\rm norm}$ or, in words, that the
state is normalized.
Unnormalized states are of the form $\lambda {\bf p} +(1-\lambda) {\bf
0}$ where $0\leq \lambda < 1$.  Unnormalized states are therefore
mixtures and hence, all pure states are normalized, that is
\[ S_{\rm pure} \subset S_{\rm norm}  \]

We define the normalization coefficient of a state ${\bf p}$ to be
\begin{equation}\label{pnormcoef}
     \mu = {\bf r}^I\cdot {\bf p}
\end{equation}
In the case where ${\bf p} \in S_{\rm norm}$ we have $\mu=1$.

The normalization coefficient is equal to the proportion of cases in
which the release button is pressed.  It is therefore a property of the
state and cannot depend on the knob setting on the measurement
apparatus.   We can see that ${\bf r}^I$ must be unique since if there
was another such vector satisfying (\ref{pnormcoef}) then this would
reduce the number of parameters required to specify the state
contradicting our starting point that a state is specified by $K$
real numbers. Hence ${\bf r}^I$ is independent of the measurement apparatus
knob setting.

\subsection{Basis states}

Any physical system can be in various states.  We expect there to exist
some sets of normalized states which are distinguishable from one
another in a single shot measurement (were this not the case then we
could store fixed records of information in such physical systems).
For such a
set we will have a setting of the knob on the measurement apparatus such
that each state in the set always gives rise to a particular outcome or
set of outcomes which is disjoint from the outcomes associated with the other
states.  It is possible that there are some non-null outcomes of the
measurement that are not activated by any of these states.  Any such
outcomes can be added to the set of outcomes associated with, say, the
first member of the set without effecting the property that the states
can be distinguished.
Hence, if these states are ${\bf p}_n$ and the measurements that distinguish
them are ${\bf r}_n$ then we have
\begin{equation}
{\bf r}_m \cdot {\bf p}_n =\delta_{mn} \quad {\rm where}\quad
\sum_{n}{\bf r}_n = {\bf r}^I
\end{equation}
The measurement vectors ${\bf r}_n$ must add to ${\bf r}^I$ since they
cover all possible outcomes.  There may be
many such sets having different numbers of elements.  Let $N$ be the
maximum number of states in any such set of distinguishable states.  We
will call $N$ the {\it dimension}.  We will call the states ${\bf p}_n$
in any such set {\it basis states} and we will call the corresponding
measurements ${\bf r}_n$ {\it basis measurements}. Each type of physical
system will be characterized by $N$ and $K$.
A note on notation: In general we will adopt the convention that the
subscript $n$ ($n=1$ to $N$) labels basis states and measurements and
the superscript
$k$ ($k=1$ to $K$) labels fiducial measurements and (to be introduced
later) fiducial states.  Also, when we need to work
with a particular choice of fiducial measurements (or states) we will take the
first $n$ of them to be equal to a basis set. Thus, ${\bf
r}^k={\bf r}_k$ for $k=1$ to $N$.

If a particular basis state is impure then we can always replace it with
a pure state.  To prove this we note that if the basis state is impure
we can write it as a convex sum of pure states. If the basis state is
replaced by any of the states in this convex sum this must also satisfy
the basis property.  Hence, we can always choose our basis sets to
consist only of pure states and we will assume that this has been done
in what follows.

Note that $N=1$ is the smallest value $N$ can take since we can always
choose any normalized state as ${\bf p}_1$ and ${\bf r}_1={\bf r}^I$.

\subsection{Simplicity}

There will be many different systems having different $K$ and $N$.  We
will assume that, nevertheless, there is a certain constancy in nature
such that $K$ is a function of $N$.  The second axiom is
\begin{axiom}
{\rm Simplicity}. $K$ is determined by a function of $N$ (i.e.
$K=K(N)$) where $N=1,2,\dots$ and where, for any given $N$,
$K$ takes the minimum value consistent with the axioms.
\end{axiom}
The assumption that $N=1,2,\dots$ means that we assume nature
provides systems of all different dimensions.
The motivation for taking the smallest value of $K$ for each given $N$
is that this way we end up with the simplest theory consistent with
these natural axioms. It will be shown that the axioms imply
$K=N^r$ where $r$ is an integer. Axiom 2 then dictates that we should
take the smallest value of $r$ consistent with the axioms (namely $r=2$).
However, it would be interesting either to show that higher values of
$r$ are inconsistent with the axioms even without this constraint that
$K$ should take the minimum value, or to explicitly construct theories
having higher values of $r$ and investigate their properties.

\subsection{Subspaces}

Consider a basis measurement set ${\bf r}_n$. The states in a basis are
labeled by the integers $n=1$ to $N$. Consider a subset $W$ of these
integers. We define
\begin{equation}
{\bf r}^{I_W}=\sum_{n\in W} {\bf r}_n
\end{equation}
Corresponding to the subset $W$ is a subspace which we will also call
$W$ defined by
\begin{equation}
{\bf p}\in W \qquad {\rm iff}\qquad
{\bf r}^{I_W}\cdot{\bf p}={\bf r}^{I}\cdot{\bf p}
\end{equation}
Thus, ${\bf p}$ belongs to the subspace if it has support only in the
subspace.  The dimension of the subspace $W$ is equal to the number of
members of the set $W$. The complement subset $\overline{W}$ consists of
the the integers $n=1$ to $N$ not in $W$.  Corresponding to the subset
$\overline{W}$ is the subspace $\overline{W}$ which we will call the
complement subspace to $W$.
Note that this is a slightly unusual usage of the
terminology ``subspace'' and ``dimension''
which we employ here because of the analogous concepts in quantum theory.
The third axiom concerns such subspaces.
\begin{axiom}
{\rm Subspaces}. A system whose state is constrained to belong to an $M$
dimensional subspace behaves like a system of dimension $M$.
\end{axiom}
This axiom is motivated by the intuition that any
collection of distinguishable states should be on an equal
footing with any other collection of the same number distinguishable
states.  In logical terms, we can think of distinguishable states as
corresponding to a propositions.  We expect a probability theory
pertaining to $M$ propositions to be independent of whether these
propositions are a subset or some larger set or not.

One application of the subspace axiom which we will use is the
following: If a system is prepared in a state which is constrained to a
certain subspace $W$ having dimension $N_W$ and a measurement is made
which may not pertain to
this subspace then this measurement must be equivalent (so far as
measured probabilities on states in $W$ are concerned) to some
measurement in the set of allowed measurements for a system actually
having dimension $N_W$.

\subsection{Composite systems}\label{compositesystems}

It often happens that a preparation device ejects its system in such a
way that it can be regarded as being made up of two subsystems.  For
example, it may emit one system to the left and one to the right (see
Fig. 2).  We will label these subsystems $A$ and $B$.  We assume
\begin{axiom} {\rm Composite systems.}  A composite system consisting of two
subsystems $A$ and $B$ having dimension $N_A$ and $N_B$
respectively, and number of degrees of freedom $K_A$ and $K_B$
respectively, has dimension $N=N_AN_B$ and number of degrees of freedom
$K=K_AK_B$.
\end{axiom}
We expect that $N=N_AN_B$ for the following reasons. If
subsystems $A$ and $B$ have $N_A$ and $N_B$ distinguishable states, then
there must certainly exist $N_AN_B$ distinguishable states for the
whole system.  It is possible that there exist more than this but we
assume that this is not so.
We will show that the relationship $K=K_AK_B$ follows from the following
two assumptions
\begin{itemize}
\item If a subsystem is in a pure state then any joint probabilities
between that subsystem and any other subsystem will factorize.  This is
a reasonable assumption given the intuition (mentioned earlier)
that pure states represent definite states for a system and therefore
should not be correlated with anything else.
\item The number of degrees of freedom associated with the full class of
states for the composite system is not greater than the number of
degrees of freedom associated with the separable states.  This is
reasonable since we do not expect there to be more entanglement than
necessary.
\end{itemize}
Note that although these two assumptions motivate the relationship
$K=K_AK_B$ we do not actually need to make them part of our axiom set
(rather they follow from the five axioms).
To show that these assumptions imply $K=K_AK_B$ consider
performing the $i$th
fiducial measurement on system $A$ and the $j$th fiducial measurement on
system $B$ and measuring the joint probability $p_{ij}$ that both
measurements have a positive outcome. These joint probabilities can be
arranged in a matrix $\tilde{p}_{AB}$ having entries $p_{ij}$.
It must be possible to choose $K_A$ linearly independent pure states
labeled ${\bf p}^{k_A}_A$ ($k_A=1$ to $K_A$) for subsystem $A$, and
similarly for subsystem $B$. With the first assumption above we can write
$\tilde{p}^{k_A k_B}_{AB}={\bf p}^{k_A}_A({\bf p}_B^{k_B})^T$ when
system $A$ is prepared in the pure state ${\bf p}^{k_A}_A$ and system $B$ is
prepared in the pure state ${\bf p}^{k_B}_B$.  It is easily
shown that it follows
from the fact that the states for the subsystems are linearly
independent that the $K_AK_B$ matrices $\tilde{p}_{AB}^{k_Ak_B}$ are linearly
independent. Hence, the vectors describing the corresponding joint
states are linearly independent.  The convex hull of the end points of
$K_AK_B$ linearly
independent vectors and the null vector is $K_AK_B$ dimensional. We
cannot prepare any additional `product' states which are linearly
independent of these since the subsystems are spanned by the set of
fiducial states considered.
Therefore, to describe convex combinations of the separable states
requires $K_AK_B$ degrees of freedom and hence, given the second
assumption above, $K=K_AK_B$.

It should be emphasized that it is not required by the axioms that the
state of a composite system should be in the convex hull of
the product states. Indeed, it
is the fact that there can exist vectors not of this form that leads to
quantum entanglement.

\section{The continuity axiom}\label{axiomfive}

Now we introduce the axiom which will give us quantum theory rather than
classical probability theory.
Given the intuition that pure states represent definite states of a
system we expect to be able to transform the state of a system
from any pure state to any other pure state.  It should be possible to
do this in a way that does not extract information about the state and
so we expect this can be done by a reversible transformation.
By reversible we mean that the effect of the transforming device
(the middle box in Fig.\ 1.) can be reversed irrespective of the input
state and hence that $Z^{-1}$ exists and is in $\Gamma$.
Furthermore, we expect any such transformation to be continuous since
there are generally no discontinuities in physics.
These considerations motivate the next axiom.
\begin{axiom}
{\rm Continuity}.
There exists a continuous reversible transformation on a system between
any two pure states of the system.
\end{axiom}
By a continuous transformation we mean that one which can be made up from many
small transformations only infinitesimally different from the identity.
The set of reversible transformations will form a compact Lie group
(compact because its action leaves the components of ${\bf p}$ bounded by
0 and 1 and hence the elements of the transformation matrices $Z$ must
be bounded).

If a reversible transformation
is applied to a pure state it must necessarily output a pure state.  To
prove this assume the contrary.  Thus, assume $Z{\bf p}=\lambda{\bf
p}_A+(1-\lambda){\bf p}_B$ where ${\bf p}$ is pure, $Z^{-1}$ exists and
is in $\Gamma$, $0<\lambda<1$, and the states ${\bf p}_{A,B}$ are distinct.
It follows that ${\bf p}=\lambda Z^{-1}{\bf p}_A+(1-\lambda)Z^{-1}{\bf p}_B$
which is a mixture. Hence we establish proof by contradiction.

The infinitesimal transformations which make up a reversible
transformation must themselves be reversible. 
Since reversible transformations always transform pure states to pure
states it follows from this axiom that we can transform any pure state
to any other pure state along a continuous trajectory through the pure states.
We can see immediately that classical systems of finite
dimension $N$ will run into problems with the continuity part of this
axiom since there are only $N$ pure states for such systems and hence
there cannot exist a continuous trajectory through the pure states.
Consider, for example, transforming a
classical bit from the state $0$ to the state $1$.  Any continuous
transformation would have to go through an infinite number of other pure
states (not part of the subspace associated with our system).
Indeed, this is clear given any
physical implementation of a classical bit.  For example, a ball in one
of two boxes must move along a continuous path from one box (representing
a 0) to the other box (representing a 1).  Deutsch has pointed out that
for this reason, the classical description is necessarily approximate in
such situations whereas the quantum description in the analogous
situation is not approximate \cite{deutsch}.  We will use this axiom to
rule out various theories which do not correspond to quantum theory
(including classical probability theory).

Axiom 5 can be further motivated by thinking about computers.  A
classical computer will only employ a finite number of distinguishable
states (usually referred to as the memory of the computer - for example
10Gbytes).  For this reason it is normally said that the computer operates
with finite resources. However, if we
demand that these bits are described classically and that
transformations are continuous then we have to invoke the existence of
a continuous infinity of distinguishable states not in the subspace
being considered.  Hence, the resources used by a classically described
computer performing a finite calculation must be infinite.  It would seem
extravagant of nature to employ infinite resources in performing a
finite calculation.

\section{The Main Proofs}

In this section we will derive quantum theory and, as an aside,
classical probability theory by dropping Axiom 5.
The following proofs lead to quantum theory

\begin{enumerate}
\item Proof that $K=N^r$ where $r=1,2,\dots$.
\item Proof that a valid choice of fiducial measurements is where we
choose the first $N$ to be some basis set of measurements and then we
choose 2 additional measurements in each of
the ${1\over2}N(N-1)$ two-dimensional subspaces (making a total of
$N^2$).
\item Proof that the state can be represented by an ${\bf r}$-type
vector.
\item Proof that pure states must satisfy an equation ${\bf r}^T D{\bf
r}=1$ where $D=D^T$.
\item Proof that $K=N$ is ruled out by Axiom 5 (though
leads to classical probability theory if we drop Axiom 5) and hence that
$K=N^2$ by the Axiom 2.
\item We show that the $N=2$ case corresponds to the Bloch sphere and
hence we obtain quantum theory for the $N=2$ case.
\item We obtain the trace formula and the conditions imposed by quantum
theory on $\hat{\rho}$ and $\hat{A}$ for general $N$.
\item We show that the most general evolution consistent with the axioms
is that of quantum theory and that the tensor product structure is
appropriate for describing composite systems.
\item We show that the most general evolution of the state after
measurement is that of quantum theory (including, but not restricted to,
von Neumann projection).
\end{enumerate}

\subsection{Proof that $K=N^r$}\label{secKNr}

In this section we will see that $K=N^r$ where $r$ is a positive
integer.  It will be shown in Section \ref{notclassical} that $K=N$
(i.e. when $r=1$) is ruled out by Axiom 5. Now, as shown in Section
\ref{qtheory}, quantum theory is consistent with the Axioms and has
$K=N^2$.  Hence, by the simplicity axiom (Axiom 2), we must have $K=N^2$
(i.e. $r=2$).

It is quite easy to show that $K=N^r$.  First note that it
follows from the subspace axiom (Axiom 3) that $K(N)$ must be a strictly
increasing function of $N$.  To see this consider first an $N$
dimensional system. This will have $K(N)$ degrees of freedom. Now
consider an $N+1$ dimensional system. If the state is constrained to
belong to an $N$ dimensional subspace $W$ then it will, by Axiom 3, have
$K(N)$ degrees of freedom. If it is constrained to belong to the
complement 1 dimensional subspace then, by Axiom 3, it will have at
least one degree of freedom (since $K$ is always greater than or equal
to 1).  However, the state could also be a mixture of a state
constrained to $W$ with some weight $\lambda$ and a state constrained to
the complement one dimensional subspace with weight $1-\lambda$.  This
class of states must have at least $K(N)+1$ degrees of freedom (since
$\lambda$ can be varied).  Hence, $K(N+1)\geq K(N)+1$.  By Axiom 4 the
function $K(N)$ satisfies
\begin{equation}
K(N_AN_B)=K(N_A)K(N_B)
\end{equation}
Such functions are known in number theory as {\it completely
multiplicative}.  It is shown in Appendix 2 that all strictly
increasing completely multiplicative functions are of the form
$K=N^\alpha$.  Since $K$ must be an integer it follows that the power,
$\alpha$, must be a positive integer. Hence
\begin{equation}
K(N)=N^r {\rm ~~~~~where~~~~~} r=1,2,3,\dots
\end{equation}
In a slightly different context, Wootters has also come to this equation
as a possible relation between $K$ and $N$ \cite{wootters}.

The signatures (see Section \ref{qtheory}) associated with $K=N$ and
$K=N^2$ are ${\bf x}=(1,0,0,\dots)$ and ${\bf x}=(1,2,0,0,\dots)$ respectively.
It is interesting to consider some of those cases that have been ruled out.
Real Hilbert spaces have ${\bf x}=(1,1,0,0,\dots)$ (consider counting
the parameters in the density matrix).  In the real Hilbert space composite
systems have more degrees of freedom than the product of the number of
degrees of freedom associated with the subsystems (which implies that
there are necessarily some degrees of freedom that can only be measured
by performing a joint measurement on both subsystems).
Quaternionic Hilbert spaces have
${\bf x}=(1,4,0,0,\dots)$.  This case is ruled out because
composite systems would have to have less degrees of
freedom than the product of the number of degrees of freedom associated
with the subsystems \cite{fuchs}. This shows that quaternionic systems
violate the principle that joint probabilities factorize when one (or
both) of the subsystems is in a pure state.
We have also ruled out $K=N^3$ (which
has signature ${\bf x}=(1,6,6,0,0,\dots)$) and
higher $r$ values.  However, these cases have only been ruled out by
virtue of the fact that Axiom 2 requires we take the simplest case.  It
would be interesting to attempt to construct such higher power theories
or prove that such constructions are ruled out by the axioms even
without assuming that $K$ takes the minimum value for each given $N$.

The fact that $x_1=1$ (or, equivalently, K(1)=1) is interesting.  It
implies that if we have a
set of $N$ distinguishable basis states they must necessarily be pure.
After the one degree of freedom associated with normalization has been
counted for a one
dimensional subspace there can be no extra degrees of freedom. If the
basis state was mixed then it could be written as a convex sum of pure
states that also satisfy the basis property. Hence, any convex sum would
would satisfy the basis property and hence there would be an extra
degree of freedom.

\subsection{Choosing the fiducial measurements}\label{choosing}

We have either $K=N$ or $K=N^2$. If $K=N$ then a suitable choice of
fiducial measurements is a set of basis measurements. For the case
$K=N^2$ any set of $N^2$ fiducial measurements that correspond to linearly
independent vectors will suffice as a fiducial set.  However, one
particular choice will turn out to be especially useful.  This choice
is motivated by the fact that the signature is
${\bf x}=(1,2,0,0,\dots)$.
This suggests that we can choose the first $N$ fiducial measurements to
correspond to a particular basis set of measurements ${\bf r}_n$ (we
will call this the fiducial basis set) and that for
each of the ${1\over2}N(N-1)$ two-dimensional fiducial subspaces
$W_{mn}$ (i.e.
two-dimensional subspaces associated with the $m$th and $n$th basis
measurements) we can chose a further two fiducial measurements which we can
label ${\bf r}_{mnx}$ and ${\bf r}_{mny}$ (we are simply using $x$ and
$y$ to label these measurements). This makes a total of $N^2$ vectors.
It is shown in Appendix 3.4 that we can, indeed, choose $N^2$ linearly
independent measurements
(${\bf r}_n$, ${\bf r}_{mnx}$, and ${\bf r}_{mny}$)
in this way and, furthermore, that they have the property
\begin{equation}\label{nooverlap}
{\bf r}_{mnx} \cdot {\bf p} = 0 ~~~{\rm if}~~~ {\bf p}\in \overline{W}_{mn}
\end{equation}
where $\overline{W}_{mn}$ is the complement subspace to $W_{mn}$.
This is a useful property since it implies that the fiducial measurements
in the $W_{mn}$ subspace really do only apply to that subspace.

\subsection{Representing the state by ${\bf r}$}

Till now the state has been represented by ${\bf p}$ and a measurement
by ${\bf r}$.  However, by introducing fiducial states, we
can also represent the measurement by a ${\bf
p}$-type vector (a list of the probabilities obtained for this
measurement with each of the fiducial states) and, correspondingly, we
can describe the state by an ${\bf r}$-type vector.
For the moment we will label vectors pertaining to the state of the
system with subscript $S$ and vectors pertaining to the measurement with
subscript $M$ (we will drop these subscripts later since it will be
clear from the context which meaning is intended).

\subsubsection{Fiducial states}

We choose $K$ linearly independent states, ${\bf p}^k_S$ for $k=1$
to $K$, and call them fiducial
states (it must be possible to choose $K$ linearly independent states
since otherwise we would not need $K$ fiducial measurements to determine
the state).  Consider a given measurement ${\bf r}_M$.  We can write
\begin{equation}
p_M^k = {\bf r}_M \cdot {\bf p}^k_S
\end{equation}
Now, we can take the number $p_M^k$ to be the $k$th component of a
vector.  This vector, ${\bf p}_M$, is related to ${\bf r}_M$ by a linear
transformation.  Indeed, from the above equation we can write
\begin{equation}\label{pCr}
{\bf p}_M = C {\bf r}_M
\end{equation}
where $C$ is a $K\times K$ matrix with $l,k$ entry equal to the $l$th
component of ${\bf p}_S^k$.  Since the vectors ${\bf p}_S^k$ are linearly
independent, the matrix $C$ is invertible and so ${\bf r}_M$
can be determined from ${\bf p}_M$. This means that ${\bf p}_M$ is an
alternative way of specifying the measurement.
Since $p_{\rm meas}$ is linear in
${\bf r}_M$ which is linearly related to ${\bf p}_M$ it must
also be linear in ${\bf p}_M$.  Hence we can write
\begin{equation}\label{formpr}
  p_{meas} = {\bf p}_M \cdot {\bf r}_S
  \end{equation}
where the vector  ${\bf r}_S$ is an alternative way of describing
the state of the system.  The $k$th fiducial state can be represented by an
${\bf r}$-type vector, ${\bf r}^k_S$, and is equal to that vector which
picks out the $k$th component of ${\bf p}_M$.  Hence, the fiducial
states are
\begin{equation}\label{fidS}
{\bf r}^1_S= \left(
\begin{matrix} 1 \\ 0 \\ 0 \\ \vdots \\ 0 \end{matrix} \right) \qquad
{\bf r}^2_S= \left(
\begin{matrix} 0 \\ 1 \\ 0 \\ \vdots \\ 0 \end{matrix} \right) \qquad
{\bf r}^3_S= \left(
\begin{matrix} 0 \\ 0 \\ 1 \\ \vdots \\ 0 \end{matrix} \right) \qquad
{\rm etc.}
\end{equation}

\subsubsection{A useful bilinear form for $p_{\rm meas}$}

The expression for $p_{\rm meas}$ is linear in both
${\bf r}_M$ and ${\bf r}_S$.  In other words, it is a bilinear form and can
be written
\begin{equation}\label{formrr}
   p_{\rm meas} = {\bf r}_M^T D {\bf r}_S
\end{equation}
where superscript $T$ denotes transpose, and
$D$ is a $K\times K$ real matrix (equal, in fact, to $C^T$).
The $k,l$ element of $D$ is
equal to the probability measured when the $k$th fiducial measurement is
performed on the $l$th fiducial state (since, in the fiducial cases,
the ${\bf r}$ vectors have one 1 and otherwise 0's as components).
Hence,
\begin{equation}\label{formfid}
D_{lk}=({\bf r}^l_M)^T D {\bf r}^k_S
\end{equation}
$D$ is invertible since the fiducial set of states are linearly
independent.

\subsubsection{Vectors associated with states and measurements}

There are two ways of describing the state:  Either with a ${\bf
p}$-type vector or with an ${\bf r}$-type vector.  From (\ref{formrp},
\ref{formrr}) we see that the relation between these two types of
description is given by
\begin{equation}\label{pDr}
{\bf p}_S = D {\bf r}_S
\end{equation}
Similarly, there are two ways of describing the measurement: Either with
an ${\bf r}$-type vector or with a {\bf p}-type vector.  From
(\ref{formpr},\ref{formrr}) we see that the relation
between the two ways of describing a measurement is
\begin{equation}
{\bf p}_M = D^T {\bf r}_M
\end{equation}
(Hence, $C$ in equation (\ref{pCr}) is equal to $D^T$.)

Note that it follows from these equations that the set of
states/measurements ${\bf r}_{S,M}$ is bounded since ${\bf p}_{S,M}$ is
bounded (the entries are probabilities) and $D$ is invertible (and hence
its inverse has finite entries).

\subsection{Pure states satisfy ${\bf r}^T D {\bf r}=1$}

Let us say that a measurement {\it identifies} a state if, when that
measurement is performed on that state, we obtain probability one.
Denote the basis measurement vectors by ${\bf r}_{Mn}$ and the basis
states (which have been chosen to be pure states)
by ${\bf p}_{Sn}$ where $n=1$ to $N$. These satisfy
${\bf r}_{Mm}\cdot{\bf p}_{Sn}=\delta_{mn}$. Hence,
${\bf r}_{Mn}$ identifies ${\bf p}_{Sn}$.

Consider an apparatus set up to measure ${\bf r}_{M1}$.  We could place
a transformation device,  $T$, in front of this which performs a reversible
transformation.  We would normally say that that $T$ transforms the
state and then ${\bf r}_{M1}$ is measured.  However, we could equally
well regard the transformation device $T$ as part of the measurement
apparatus.  In this case some other measurement ${\bf r}$ is being
performed.  We will say that any measurement which can be regarded as a
measurement of ${\bf r}_{M1}$ preceded by a reversible transformation
device is a {\it pure measurement}.  It is shown in Appendix 3.7 that all
the basis measurement vectors ${\bf r}_{Mn}$ are pure measurements and,
indeed, that the set of fiducial measurements of Section \ref{choosing}
can all be chosen to be pure.

A pure measurement will
identify that pure state which is obtained by acting on ${\bf p}_{S1}$
with the inverse of $T$. Every pure state can be reached in this
way (by Axiom 5) and hence, corresponding to each pure state there
exists a pure measurement.  We show in Appendix 3.5 that the map between
the vector representing a pure state and the vector representing
the pure measurement it is identified by  is linear and invertible.

We will now see that not only is this map linear but also that, by
appropriate choice of the fiducial measurements and fiducial states, we
can make it equal to the identity.
A convex structure embedded in a $K$-dimensional space must have at
least $K+1$ extremal points (for example, a triangle has three extremal
points, a tetrahedron has four, etc.). In the case of the set $S$, one
of these extremal points will be ${\bf 0}$ leaving at least $K$
remaining extremal points which will correspond to pure states (recall
that pure states are extremal states other than ${\bf 0}$).
Furthermore, it must be possible to choose a set of $K$ of these
pure states to correspond to linearly independent vectors (if this
were not possible then the convex hull would be embedded in a lower than
$K$ dimensional space).
Hence, we can choose all our fiducial states to be pure.
Let these fiducial states be ${\bf r}^k_S$. We will choose
the $k$th fiducial measurement ${\bf r}^k_M$ to be that pure measurement which
identifies the $k$th fiducial state. These will constitute a linearly
independent set since the map from the corresponding linearly
independent set of states is invertible.

We have proven (in Appendix 3.5)
that, if ${\bf r}_M$ identifies ${\bf r}_S$, there must exist a map
\begin{equation}
{\bf r}_S = H {\bf r}_M
\end{equation}
where $H$ is a $K\times K$ constant matrix.  In particular this is true for
the fiducial states and fiducial measurements:
\begin{equation}
{\bf r}^k_S = H {\bf r}^k_M
\end{equation}
However, the fiducial vectors have the special form given in
(\ref{fidM},\ref{fidS}), namely zeros everywhere except for the $k$th
entry.  Hence, the map $H$ is equal to the identity.
This is true because we have chosen the fiducial measurements to be
those which identify the fiducial states. Since these vectors are
related by the identity map we will drop the $M$ and $S$
subscripts in what follows, it being understood that the left most
vector corresponds to the measurement apparatus and the right most vector
corresponds to the state.  {\it Thus the measurement ${\bf r}$ identifies the
state ${\bf r}$ (i.e.\ given by the same vector) if ${\bf r}$ is pure.}  Hence,
\begin{equation}
{\bf r}^T D {\bf r}=1
\end{equation}
for pure states (and pure measurements).
This equation is very useful since will help us to find the pure states.
It is shown in Appendix 3.6 that $D=D^T$.

It is shown in Appendix 3.7 that the fiducial measurements ${\bf r}_n$,
${\bf r}_{mnx}$, and ${\bf r}_{mny}$ are pure.  They will identify a set
of pure states represented by the same vectors ${\bf r}_n$,
${\bf r}_{mnx}$, and ${\bf r}_{mny}$ which we take to be our fiducial
states.  The first $N$ fiducial states, ${\bf r}_n$, are then just the
basis states and it follows from (\ref{nooverlap}) that the remaining
basis states, ${\bf r}_{mnx}$ and ${\bf r}_{mny}$, are in the
corresponding $W_{mn}$ subspaces.

\subsection{Ruling out the $K=N$ case}\label{notclassical}

Consider the $K=N$ case.  There will be $K=N$ fiducial vectors
which we can choose to be equal to the basis vectors.  From equation
(\ref{formfid}) we know that the $lk$ element of
$D$ is equal to the measured probability with the $k$th fiducial state
and the $l$th fiducial measurement.  Since the fiducial vectors
correspond to basis vectors this implies that $D$ is equal to the
identity.  The pure states must
satisfy
\begin{equation}
{\bf r}^T D {\bf r} = 1
\end{equation}
We also have ${\bf p}=D{\bf r}$ (equation (\ref{pDr})).  Given that $D$
is equal to the identity in this case we obtain
\begin{equation}\label{classical}
\sum_{k=1}^N (p^k)^2=1
\end{equation}
where ${p^k}$ is the $k$th component of ${\bf p}$.  However,
\begin{equation}\label{physprob}
0\leq p^k \leq 1
\end{equation}
Normalization implies that
\begin{equation}\label{pnormal}
\sum_{k=1}^N p^k=1
\end{equation}
The solutions of (\ref{classical}), (\ref{physprob}),
(\ref{pnormal}) have one
$p_k$ equal to 1 and all the others are equal to 0.  In other words,
the only pure vectors are the basis vectors themselves which corresponds
to classical probability theory. This forms
a discrete set of vectors and so it is impossible for Axiom 5 (the continuity
axiom) to be satisfied.  Hence, we rule out such theories.  However, if
Axiom 5 is dropped then, by Axiom 2, we must take $K=N$.  This
necessarily corresponds to classical probability theory for the
following reasons.  We can choose our $K$ ($=N$) fiducial measurements
to be the basis measurements ${\bf r}_n$.  Then the basis states must be
represented by vectors with zero's in all positions except the $n$th
position.  All states must have normalization coefficient less than or
equal to 1.  Hence, all states can be written as a convex combination of
the basis states and the null state.  This means that only the basis
states are pure states.  Hence, we have classical probability theory.

\subsection{The Bloch sphere}\label{secbloch}

We are left with $K=N^2$ (since $K=N$ has been ruled out by Axiom 5).
Consider the simplest non-trivial case $N=2$ and $K=4$.  Normalized
states are contained in a $K-1=3$ dimensional convex set.  The surface
of this set is two-dimensional.  All pure states correspond to points on
this surface.  The four fiducial states can all be taken to be pure.  They
correspond to a linearly independent set.
The reversible transformations that can act on the states form a
compact Lie Group.  The Lie dimension (number of generators) of this
group of reversible transformations cannot be equal to one since, if
it were, it could not transform between the fiducial states. This is because,
under a change
of basis, a compact Lie group can be represented by
orthogonal matrices \cite{boerner}. If there
is only one Lie generator then it will generate pure states on a circle.  But
the end points of four linearly independent vectors cannot lie on a
circle since this is embedded in a two-dimensional subspace.  Hence, the
Lie dimension must be equal to two.  The pure states are
represented by points on the two-dimensional surface.  Furthermore,
since the Lie
dimension of the group of reversible transformations is equal to two
it must be possible to
transform a given pure state to any point on this surface.  If we can
find this surface then we know the pure states for $N=2$.  This surface
must be convex since all points on it are extremal.  We will use this
property to show that the surface is ellipsoidal and that, with
appropriate choice of fiducial states, it can be made spherical (this is
the Bloch sphere).

The matrix $D$ can be calculated from equation (\ref{formfid})
\[ D_{ij}= ({\bf r}^i)^T D {\bf r}^j  \]
As above, we will choose the fiducial measurements to be those pure
measurements which identify the fiducial
states (these also being taken to be pure).  Hence, $D$ will have 1's along
the diagonal.  We choose the first two fiducial vectors to be basis
vectors.  Hence, $D$ has the form
\begin{equation}\label{twodDsym}
D=\left( \begin{array}{cccc} 1 &  0  &  1-a & 1-b \\
                             0 &  1  &   a  &  b  \\
                             1-a &  a &   1  &  c  \\
                             1-b &  b &   c &  1   \end{array} \right)
\end{equation}
The two 0's follow since the first two vectors are basis vectors (i.e.\
$({\bf r}^1)^T D {\bf r}^2=0$ and $({\bf r}^2)^T D {\bf r}^1=0$).  The
$1-a$ and $a$ pair above the diagonal follow from normalization since
\begin{equation}
1=({\bf r}^I)^T D {\bf r}^i=({\bf r}^1)^T D {\bf r}^i +
({\bf r}^2)^T D {\bf r}^i
\end{equation}
The $1-b$ and $b$ pair follow for similar reasons.  The matrix is
symmetric and this gives all the terms below the diagonal.

We will not show that the constraints on the elements of $D$ are the
same as in quantum theory (discussed in Section \ref{qtheory}).
Define
\begin{equation}\label{rtov}
{\bf v} = \left( \begin{array}{c} v_0 \\ v_1 \\ v_2 \\ v_3 \end{array}
\right) = \left( \begin{array}{c} r_1 \\ r_2-r_1 \\ r_3 \\ r_4 \end{array}
\right) \end{equation}
Thus,
\begin{equation}
{\bf r} = C {\bf v}
\end{equation}
where
\begin{equation}
         C=\left( \begin{array}{cccc} 1 & 0 & 0 & 0 \\
                                     1 & 1 & 0 & 0 \\
                                     0 & 0 & 1 & 0 \\
                                     0 & 0 & 0 & 1  \end{array}
                                     \right)
\end{equation}
Hence ${\bf r}^T D {\bf r}' = {\bf v}^T C^TDC {\bf v}'$.  From
(\ref{twodDsym}) we obtain
\begin{equation}
F\equiv C^TDC =\left( \begin{array}{cccc} 2 &  1 &  1 & 1  \\
                             1 &  1 &   a  &  b  \\
                             1&  a &  1 &  c  \\
                            1 &  b &  c&  1   \end{array} \right)
\end{equation}
Now, ${\bf r}^I={\bf r}_1+{\bf r_2}=(1, 1, 0, 0)^T$.
The corresponding ${\bf v}$ type
vector is, using (\ref{rtov}), ${\bf v}^I= (1 , 0, 0, 0)^T$.  Assume
that ${\bf r}$ is normalized to $\mu$ and ${\bf r}'$ is normalized to
$\mu'$.  Then
\begin{equation}\label{vnorm}
\mu = {\bf v}^I F {\bf v} = 2v_0 + \sum_{i=1}^3 v_i
\end{equation}
and similarly for $\mu'$.  For normalized states $\mu=1$.
If ${\bf v}^T F {\bf v}'$ is multiplied out and (\ref{vnorm}) is used to
eliminate $v_0$ (and a similar equation is used to eliminate $v_0'$) then
we obtain
\begin{equation}\label{bloch}
p_{\rm meas} = {\bf r}^T D {\bf r}' = \vec{v}^T A \vec{v}\,'
+ \mu\mu'/2
\end{equation}
where
\begin{equation}
 \vec{v}= \left( \begin{array}{c} v_1 \\ v_2 \\ v_3 \end{array}\right)
=  \left( \begin{array}{c} r_2-r_1 \\ r_3 \\ r_4
 \end{array}\right)
\end{equation}
and
\begin{equation}
A= \left( \begin{array}{ccc}  {1\over 2} &   a-{1\over 2}  &  b -{1\over 2}  \\[2pt]
                              a-{1\over 2} & {1\over 2} &  c-{1\over 2}  \\[2pt]
                              b-{1\over 2} &  c-{1\over 2}& {1\over 2}
                              \end{array} \right)
\end{equation}

All the pure states will be normalized.
Furthermore, they will satisfy ${\bf r}^TD{\bf r}=1$ or
\begin{equation}
\vec{v}^T A \vec{v} = {1\over 2}
\end{equation}
This equation defines a two dimensional surface $T$ embedded in three
dimensions.  For example, if $a=b=c={1\over 2}$ then we have a sphere of
radius 1 (this is, in fact, the Bloch sphere).  If $A$ has three positive
eigenvalues then $T$ will be an ellipsoid.  If $A$ has one or two negative
eigenvalue then $T$ will be a hyperboloid (if
$A$ has three negative eigenvalues then there cannot be any real
solutions for $\vec{v}$). An equal mixture of the two basis
states ${1\over 2} {\bf r}_1 +{1\over 2}{\bf r}_2$ corresponds to
$\vec{v}=(0,0,0)^T$.  Thus, the origin is in the set of allowed states.
An ellipsoid represents a convex surface with the origin in its
interior.  On the other hand, the curvature of a hyperboloid is such
that it cannot represent a convex surface with the origin on the
interior and so cannot represent points in the set of pure
vectors.  Thus we require that $T$ has three positive eigenvalues.   A
necessary condition for $A$ to have all
positive eigenvalues is that $\det(A) > 0$.  We have three variables
$a$, $b$ and $c$.  The condition $\det(A)=0$ is satisfied when
\begin{equation}\label{roots}
c=c_{\pm}\equiv 1-a-b+2ab \pm 2\sqrt{ab(1-a)(1-b)}
\end{equation}
Note, we get the same conditions on $c$ if we solve $\det{D}=0$.
We know the case with $a=b=c=1/2$ corresponds to a sphere.  This falls
between the two roots in equation (\ref{roots}).  The sign of the
eigenvalues cannot change unless the determinant passes through a root
as the parameters are varied.  Hence, all values of $a$, $b$, $c$
satisfying
\begin{equation}\label{abceqn}
c_{-}< c < c_{+}
\end{equation}
must correspond to three positive eigenvalues and hence to an ellipsoid.
Values outside this range correspond to some negative eigenvalues (this
can be checked by trying a few values).  Hence, (\ref{abceqn}) must be
satisfied. This agrees with quantum theory (see (\ref{cpccm})).
Therefore, we have obtained
quantum theory from the axioms for the special case $N=2$.
As detailed in Section \ref{qtheory}, if we are given $D$ we can go back to the
usual quantum formalism by using $D$ to calculate $\hat{\bf P}$ (making
some arbitrary choices of phases) and then using the formulae in that
section (equations (\ref{ArP}) and (\ref{rhoPr}))
to obtain $\hat{\rho}$ for the state and $\hat{A}$ for the
measurement.

If $T$ is ellipsoidal it is because we have made a
particular choice of fiducial projectors $\hat{P}_k$.  We can choose a
different set to make $T$ spherical.  Since the choice of fiducial
vectors is arbitrary we can, without any loss of generality, always
take $T$ to be spherical with $a=b=c=1/2$.  Hence, without loss of
generality, we can always put
\begin{equation}\label{Dhalfs}
D= \left( \begin{array}{cccc} 1 &  0 &  \frac{1}{2} & \frac{1}{2}\\[3pt]
                             0 &  1 &   \frac{1}{2}  &  \frac{1}{2} \\[3pt]
                             \frac{1}{2} & \frac{1}{2} &  1 & \frac{1}{2}\\[3pt]
                            \frac{1}{2} & \frac{1}{2} & \frac{1}{2} &  1
                            \end{array} \right)
\end{equation}
for the $N=2$ case.

Since we have now reproduced quantum theory for the $N=2$ case we can
say that
\begin{itemize}
\item Pure states can be represented by $|\psi\rangle\langle\psi |$
where $|\psi\rangle= u|1\rangle+v|2\rangle$ and where $u$ and $v$ are
complex numbers satisfying $|u|^2+|v|^2=1$.
\item The reversible transformations which can transform one pure state
to another can be seen as rotations of the Bloch sphere, or as the
effect of a unitary operator $\hat U$ in $SU(2)$.
\end{itemize}
This second observation will be especially useful when we generalize to
any $N$.

\subsection{General $N$}

It quite easy now to use the $N=2$ result to construct the case for
general $N$ using Axiom 3 (the subspace axiom).  We will use the
$N=3$ case to illustrate this process.  For this case $K=9$ and so we
need 9 fiducial vectors which we will choose as in Section
\ref{choosing}.  Thus, we choose the first 3 of these
to be the fiducial basis vectors.
There are 3 two-dimensional fiducial
subspaces.  Each of these must have a further two fiducial vectors (in
addition to the basis vectors already counted).  As in Section
\ref{choosing} we will label the two
fiducial vectors in the $mn$ subspace as $mnx$ and $mny$.
We will choose the following order for the fiducial states
\[ 1,~ 2,~ 3,~ 12x,~ 12y,~ 13x,~ 13y,~ 23x,~ 23y    \]
This provides the required 9 fiducial vectors. These fiducial vectors
can represent pure states or pure measurements.  The matrix $D$ is a
$9\times 9$ matrix.  However, each two-dimensional fiducial subspace must, by
Axiom 3, behave as a system of dimension $2$.  Hence, if we take those
elements of $D$ which correspond to an $N=2$ fiducial subspace they must
have the form given in equation (\ref{Dhalfs}).  We can then calculate
that for $N=3$
\[ D= \left( \begin{array}{ccccccccc}
1 & 0 & 0 & h & h & h & h & 0 & 0 \\
0 & 1 & 0 & h & h & 0 & 0 & h & h \\
0 & 0 & 1 & 0 & 0 & h & h & h & h \\
h & h & 0 & 1 & h & q & q & q & q \\
h & h & 0 & h & 1 & q & q & q & q \\
h & 0 & h & q & q & 1 & h & q & q \\
h & 0 & h & q & q & h & 1 & q & q \\
0 & h & h & q & q & q & q & 1 & h \\
0 & h & h & q & q & q & q & h & 1
\end{array}\right)  \]
where $h=1/2$ and, as we will show, $q=1/4$.  All the 0's are because
the corresponding subspaces do not overlap (we are using property
(\ref{nooverlap})).  The $q$'s correspond to
overlapping subspaces.  Consider for example, the $D_{46}$ term.  This
is given by ${\bf r}_{12x}^T D {\bf r}_{13x}$ which is the probability
when ${\bf r}_{12x}$ is measured on the state ${\bf r}_{13x}$.  If
states are restricted to the $13$ fiducial subspace then, by Axiom 3,
the system must
behave as a two-dimensional system.  In this case, the measurement ${\bf
r}_{12x}$ corresponds to some measurement in the $13$ fiducial subspace.
Since it has support of $1/2$ on the $1$ basis state and support of $0$
on the $3$ basis state this measurement must be equivalent to the
measurement ${1\over 2}{\bf r}_1$ (though only for states restricted to
the $13$ fiducial subspace).  But ${\bf r}_1^T D{\bf r}_{13x}=1/2$ and
hence ${\bf r}_{12x}^T D {\bf r}_{13x}=1/4$.
We can use a similar procedure to calculate $D$ for any $N$.  Once we
have this matrix we can convert to the usual quantum formalism as we did
in the $N=2$ case.  The projection operators which give rise to this $D$
are, up to arbitrary choices in phase, those in equations (\ref{nbasis})
and (\ref{mpmn}) (these arbitrary choices in phase correspond to fixing
the gauge).  Hence, we obtain $\hat{\bf P}$.
Using the results of Section \ref{qtheory}, we obtain
\begin{equation}\label{convertstate}
\hat{\rho}=\hat{\bf P}\cdot{\bf r}
\end{equation}
for a state represented by $\bf r$, and
\begin{equation}\label{convertmeas}
\hat{A}={\bf r}\cdot\hat{\bf P}
\end{equation}
for a measurement represented by $\bf r$.  Hence, we obtain
\begin{equation}
p_{\rm meas}={\rm trace}(\hat{A}\hat{\rho})
\end{equation}
which is shown to be equivalent to $p_{\rm meas}={\bf r}\cdot{\bf p}$ in
section \ref{qtheory}. We now need
to prove that the restrictions from quantum theory on $\hat{A}$ and
$\hat{\rho}$ follow from the axioms.

Both $\hat{\rho}$ and $\hat{A}$ must be Hermitean since $\bf r$ is real.
The basis state ${\bf r}_1$ is represented by
$|1\rangle\langle 1|$.  We showed above that we can apply any unitary
rotation $U\in SU(2)$ for the $N=2$ case. It follows from Axiom 3 and
the results of the previous section that
if we apply an reversible transformation in a two-dimensional fiducial
subspace on a state which is in that two-dimensional
subspace the effect will be given by the action of a unitary operator
acting in that subspace.
Thus imagine we prepare the state $|1\rangle\langle 1|$.  Let the basis
states be $|n\rangle\langle n|$ (where $n=1$ to $N$). Perform the
rotation $U_{12}$ in the 12 subspace.  This transforms the state to
$U_{12}|1\rangle\langle 1 |U_{12}^\dagger$. Now redefine the basis
states to be $|1'\rangle\langle 1'|\equiv
U_{12}|1\rangle\langle 1 |U_{12}^\dagger$,
$|2'\rangle\langle 2'|\equiv
U_{12}|2\rangle\langle 2 |U_{12}^\dagger$, and
$|n\rangle\langle n|$ for $n\not=1,2$ (it is shown in Appendix 3.3 that
a reversible transformation in a subspace can be chosen to leave basis
states not in that subspace unchanged).  Next, we consider a rotation
$U_{1'3}$ in the 1'3 subspace.  The state will only have support in this
subspace and so Axiom 3 can be applied again.  The basis states can be
redefined again.  This process can be repeated. In this way it is easy
to prove we can generate any state of the form
\begin{equation}
\hat{\rho}=|\Psi\rangle\langle\Psi |
\end{equation}
where
\begin{equation}
|\Psi\rangle=\sum_{n=1}^N c_n|n\rangle
\end{equation}
and $\sum_n |c_n|^2=1$ (this is most easily proven by starting with the
target state and working backwards).
These transformations are reversible and hence all
the states generated in this way must be pure.  Now, since we have shown
that these states exist, all measurements performed on these states must
be non-negative.  That is
\begin{equation}
{\rm trace}(\hat{A} |\Psi\rangle\langle\Psi | ) \geq 0 ~~{\rm for ~ all}~~
|\Psi\rangle
\end{equation}
Hence, we obtain the positivity condition for the operators $\hat{A}$
associated with measurements.
For each state, ${\bf r}$, there exists a pure measurement represented
by the same
vector, ${\bf r}$, which identifies the state.  Hence, since the state
$|\Psi\rangle\langle\Psi |$
exists, it follows from (\ref{convertstate},\ref{convertmeas}) that
measurements of the form
\begin{equation}
\hat{A}=|\Psi\rangle\langle\Psi |
\end{equation}
exist.  Therefore, all states $\hat{\rho}$ must satisfy
\begin{equation}
{\rm trace}(|\Psi\rangle\langle\Psi |\hat{\rho} ) \geq 0 ~~{\rm for ~ all}~~
|\Psi\rangle
\end{equation}
Hence we have proved the positivity condition for states.

We have $\hat{I}={\bf r}^I\cdot\hat{\bf P}$ since the first $N$ elements
of ${\bf r}^I$ are equal to 1 and the remainder are 0, and the first $N$
elements of $\hat{\bf P}$ are projectors corresponding to a basis.
Hence, the trace condition (that $0\leq{\rm trace} (\hat{\rho})\leq 1$)
follows simply from the requirement $0\leq{\bf r}^I\cdot{\bf p}\leq 1$.

The most general measurement consistent with the axioms can be shown to
be a POVM.  A set of measurements ${\bf r}_l$ that can be performed with
a given knob setting on the measurement apparatus must satisfy
$\sum_l {\bf r}_l = {\bf r}^I$.  Using (\ref{convertmeas}), this
corresponds to the constraint that $\sum_l \hat{A}_l = I$ as required.

\subsection{Transformations}

It was shown in Section \ref{qtheory}
that the transformation $Z$ on ${\bf p}$ is
equivalent to the transformation $\$ $ on $\hat{\rho}$ where
\begin{equation}\label{transZprho}
Z={\rm tr}(\hat{\bf P}\$(\hat{\bf P})^T) D^{-1}
\end{equation}
To discuss the constraints on transformations we need to
consider composite systems.
Fig. 2. shows a preparation apparatus producing a system made
up of subsystems $A$ and $B$ such that $A$ goes to the left and $B$ goes
to the right.
\begin{figure*}[t]
{\includegraphics{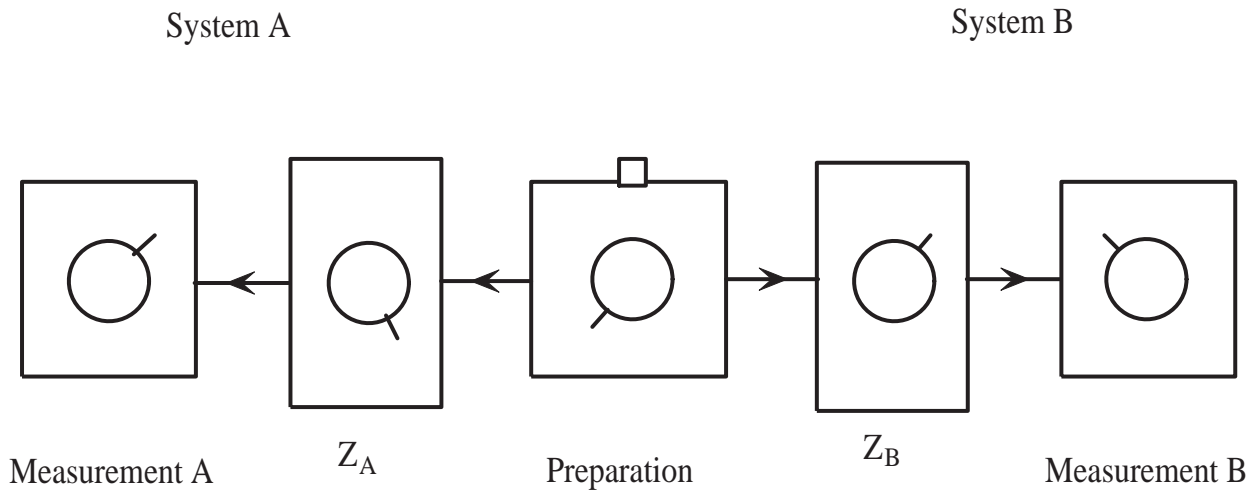}}
\caption{The preparation device here prepares a system in the form of
two subsystems which go to the left and the right.}
\end{figure*}
These subsystems then impinge on measurement
apparatuses after passing through transformations devices which perform
transformations $Z_A$ and $Z_B$.  This set up
can be understood to be a special case of the more generic setup shown
in Fig. 1. (there is no stipulation in the case of Fig. 1. that the
measurement apparatus or any of the other apparatuses be located only
in one place).  Assume the
transformation devices are initially set to leave the subsystems
unchanged.  From Axiom 4 we know that there are $K_AK_B$ fiducial
measurements. As discussed in Section \ref{qtheory}, the space of
positive operators for the composite system is spanned by
$\hat{P}^A_i\otimes\hat{P}^B_j$ where $\hat{P}^A_i$ ($i=1$ to $K_A$) is
a fiducial set for $A$ and $\hat{P}^B_j$ ($j=1$ to $K_B$)
is a fiducial set for $B$. It is shown in Appendix 4 that (as we would
expect) the projector $\hat{P}^A_i\otimes\hat{P}^B_j$ corresponds
(i) to preparing the $i$th fiducial state at side $A$ and the $j$th
fiducial state at side $B$ when the operator is regarded as representing a
state, and
(ii) to measuring the joint probability of obtaining a positive outcome
at both ends when the $i$th fiducial measurement is performed at side
$A$ and the $j$th fiducial measurement is performed at side $B$ when the
operator is regarded as representing a measurement.
Hence, one choice of fiducial measurements is where we simply
perform the $i$th fiducial measurement on $A$ and the $j$th fiducial
measurement on $B$ and measure the joint probability $p_{ij}$.
The probabilities $p_{ij}$ could be put in the form
of a column vector ${\bf p}_{AB}$.  However, for discussing transformations,
it is more convenient to put them in the form of a $K_A \times K_B$
matrix, $\tilde{p}_{AB}$, having $ij$ entry $p_{ij}$. It is easy to
convert between these two ways of describing the state.  We could
regard both the preparation apparatus and measurement apparatus B as a
preparation apparatus preparing states of subsystem $A$.  If we perform
the $j$th fiducial measurement on system $B$ and take only those cases
where we obtain a positive result for this measurement preparing the
null state otherwise then the
resulting state of system $A$ will be given by a vector equal to
the $j$th column of $\tilde{p}_{AB}$ (since these probabilities are
equal to the probabilities that would be obtained for the
fiducial measurements on $A$ with this preparation).  Hence, the columns
of $\tilde{p}_{AB}$ must transform under $Z_A$.
Similarly, the rows of $\tilde{p}_{AB}$ must
transform under $Z_B$.  Hence, when the transformation devices in Fig.
2. are active, we have
\begin{equation}\label{ZpZ}
\tilde{p}_{AB} \rightarrow Z_A \tilde{p}_{AB} Z_B^T
\end{equation}
If the state is represented by $\tilde{r}_{AB}$ where
\begin{equation}
\tilde{p}_{AB}=D_A\tilde{r}_{AB}D_B^T
\end{equation}
then this equation
becomes
\begin{equation}\label{XrX}
\tilde{r}_{AB} \rightarrow X_A \tilde{r}_{AB} X_B^T
\end{equation}
where
\begin{equation}
X_A=D^{-1}_AZ_AD_A
\end{equation}
and similarly for $B$.
It is easy to see that this is the correct transformation equation in
quantum theory (we have dropped the $A$ and $B$ superscripts).
\begin{equation}\label{tensor}
\begin{array}{rcl}
p^{ij}_{AB} &\!\!\! \rightarrow & {\rm tr}
[\hat{P}_i\otimes\hat{P}_j \$_A\otimes\$_B (\hat{\rho})  ] \\
 {} & = & {\rm tr}[\hat{P}_i\otimes\hat{P}_j\$_A\otimes\$_B
 (\sum_{kl}\hat{P}_k\otimes\hat{P}_l r^{kl} )] \\
 { } & = & \sum_{kl}{\rm tr}[\hat{P}_i\otimes\hat{P}_j
 \$_A(\hat{P}_k)\otimes\$_B(\hat{P}_l)] r^{kl} \\
 {} & = &  \sum_{kl} {\rm tr}[\hat{P}_i\$_A(\hat{P}_k)]
 r_{AB}^{kl} {\rm tr} [\hat{P}_i\$_B(\hat{P}_l)]
 \end{array}
\end{equation}
which, using (\ref{transZprho}), gives (\ref{ZpZ}) and (\ref{XrX}).  The steps
in (\ref{tensor}) can
be read backwards.  Hence, from (\ref{ZpZ}), we obtain the
tensor product structure for describing composite systems.

We will say that $Z_A$ is completely positive iff
\begin{equation}\label{ZPZ}
\tilde{p}_{AB} \rightarrow Z_A \tilde{p}_{AB}
\end{equation}
maps all allowed states of the composite system $AB$ to states
which are also allowed states for any dimension $N_B$.
The only constraint on transformation matrices $Z$ is that they
transform states in $S$ to states in $S$.  This means that probabilities
must remain bounded by 0 and 1.  Hence,
\begin{enumerate}
\item $Z$ must not increase the normalization coefficient of states.
\item $Z$ must be completely positive.
\end{enumerate}
Condition 2 is necessary since any system could always be a subsystem of
some larger system. The transformations deduced from the axioms are
subject to the equivalent constraints for $\$ $ listed in Section
\ref{qtheory}.  They preserve Hermitivity since the transformation
matrix $Z$ is real (and hence ${\bf p}$ remains real).
They do not increase the trace (point 1. above).
They are linear and they must be completely positive (point 2. above).
Hence, the most general type of transformation consistent with the
axioms is the most general transformation of quantum theory.  As noted
in section \ref{qtheory}, this gives us unitary evolution and von
Neumann projection as special cases.

\subsection{The state after a measurement}

It is possible that, after a measurement, a quantum system emerges from
the measurement apparatus.  In such cases the measurement apparatus is
also behaving as a transformation apparatus. We can think of the state
as emerging into a different channel for each measurement outcome.
Associated with each
outcome, $l$, of the measurement will be a certain transformation,
$Z_l\in \Gamma$, on the state.
The probability of any given outcome will not, in
general, be equal to 1.  Hence, the transformation must reduce the
normalization coefficient associated with the state to a value
consistent with the probability of obtaining that outcome.  This
condition is
\begin{equation}\label{meascona}
{\bf r}^I\cdot Z_l {\bf p} = {\bf r}_l\cdot{\bf p} ~~~~{\rm for~all~}
{\bf p}\in S
\end{equation}
Furthermore, we can consider all these channels taken together.  In this
case the effective transformation is given by $\sum_l Z_l$.  It is
necessary that this also belongs to the allowed set of transformations,
$\Gamma$, and that it does not change the normalization coefficient
associated with the state.  This second condition can be written
\begin{equation}\label{measconb}
\bigg(\sum_l Z_l\bigg)^T {\bf r}^I ={\bf r}^I
\end{equation}
This is equivalent to constraint
\begin{equation}
{\rm tr}\sum_l \$ (\hat{\rho}) = {\rm tr}(\hat{\rho}) ~~~{\rm
for~all}~~~\hat{\rho}
\end{equation}
Since completely positive operators can be written as
$\$(\hat{\rho})=\sum_l \hat{M}_l\hat{\rho}\hat{M}_l^\dagger $ this
equation can be shown to be equivalent to
\begin{equation}
\sum_l \hat{M}_l^\dagger \hat{M}_l = \hat{I}
\end{equation}
which is the usual quantum constraint on superoperators associated with
measurements \cite{krauss,Nielsenchuang}.

The two equations (\ref{meascona},\ref{measconb}) which constrain the
possible transformations of the state after measurement apply equally
well to classical probability theory.  This may suggest a new approach
to the measurement problem in quantum theory.

\section{Infinite dimensional spaces}

There are two types of infinite dimensional space - countable and
continuous dimensional.  The countable infinite dimensional spaces are
accounted for by these axioms since
such systems are characterized by the property that any finite subspace
obeys quantum theory.
It is not so clear what the status of continuous dimensional spaces is.
Such spaces can always be modeled arbitrarily well by a countable
infinite dimensional Hilbert space.  However, there are certain
mathematical subtleties associated with the continuous case which we
have not considered here.  Nevertheless, it is clear that the classical
continuous
case violates the axioms even though there are continuous paths between
states since the continuity axiom (Axiom 5) must also apply to finite
subspaces (by Axiom 3) and for these there are no continuous
transformations.

While continuous dimensional spaces play a role in some
applications of quantum theory it is worth asking whether we expect
continuous dimensional spaces to appear in a truly fundamental physical
theory of nature.  Considerations from quantum gravity suggest that
space is not continuous at the planck scale and that the amount of
information inside any finite volume is finite implying that the
number of distinguishable states is countable.  Given the mathematical
difficulties that appear with continuous dimensional Hilbert spaces it
is also natural to ask what our motivation for considering such spaces
was in the first place.  Consider a classical particle which can move along a
straight line.  If where were not a continuous infinity of distinguishable
positions for the particle then the only way the particle could move
would be to jump from one position to the next. It is because we do not
like such discontinuities in physics that we imagine that
there is a continuous infinity of distinct positions along the line.
However, in
quantum theory it is no longer the case that the particle would need to
jump and hence the main motivation for considering the continuous
dimensional case no longer pertains.

If we do, nevertheless, consider continuous dimensional spaces then there
is an interesting respect in which the quantum case is superior to the
classical case.  Consider again a particle which can move along a
straight line. Every point on the line represents a distinguishable
state for the particle.  Take three points $A$, $B$, and $C$ along this
line where $B$ is between $A$ and $C$.  In classical theory, if the
particle is to move continuously through the state space from $A$ to $C$
it must pass through point $B$.  However, to move continuously from $A$
to $B$ it need not pass through $C$.  Hence, the pairs $AB$ and $AC$ are
on an unequal footing.  In quantum theory a particle can pass directly
from the point $A$ to the point $C$ without going through the points in
between along a continuous trajectory in the state space simply by
going along the Bloch sphere corresponding to this two-dimensional
subspace (such transformations do not occur in practise since
Hamiltonians contain only local terms).  Hence, the pairs $AB$ and $AC$
are on an equal footing. We can regard statements like \lq\lq the
particle is at point $B$\rq\rq\,\,  as logical propositions. It is
a very desirable property that pairs of propositions should be on an
equal footing.  Thus, in this respect, quantum theory is superior.

On the other hand, even in the quantum case,
continuous dimensional spaces appear to have a topological relationship
between infinitesimally displaced distinguishable states which is
different to the topological relationship between finitely displaced
distinguishable states. This is hard to reconcile with the notion that
any pair of distinguishable states are on an equal footing and may be
further support for the case against giving continuous dimensional
spaces a role in any fundamental theory of nature.

\section{Discussion}

We have shown that quantum theory follows from five very natural axioms.
If Axiom 5 (or even just the word ``continuous'' in Axiom 5)
is dropped we obtain classical probability theory instead.  It is classical
probability theory that must have \lq jumps\rq.  If a 19th century ancestor
of Schroedinger had complained about ``dammed classical jumps''
then he might have attempted to derive a continuous theory of
probability and arrived at quantum theory.  Quantum theory is, in some
respects, both superior to and more natural than classical probability
theory (and
therefore classical theories in general) since it can describe evolution for
finite systems in a continuous way.  Since nature is quantum,
not classical, it is to be expected that quantum theory
is ultimately the more reasonable theory.

There are many reasons to look for better axiomatic formulations of
quantum theory.
\begin{itemize}
\item Aesthetics.  A theory based on reasonable axioms is more
appealing.
\item A set of reasonable axioms provides us with a deeper conceptual
understanding of a theory and is therefore more likely to suggest ways
in which we could extend the domain of the theory or modify the axioms
in the hope of going beyond quantum theory (for example, to develop
quantum gravity).
\item This approach puts a different slant on the interpretation
of quantum theory (see discussion below).
\item  Since the formulation of quantum theory here is closer to
classical probability theory than the standard formulation, this may
motivate new applications and new treatments of the theory of quantum
information.
\end{itemize}

There are various ways in which this work has a bearing on
interpretational matters.  First, if we really believe these axioms to
be reasonable then they would also apply to hidden variables and it
would follow that the hidden variable substructure must look
like quantum theory.  We could not then use hidden variables to solve
the measurement problem (since this relies on being able to give the
hidden variables a classical probability interpretation).
Second, we see here how successful a purely instrumentalist approach is
in obtaining the structure of quantum theory. Whilst this need not contradict
beliefs held by the realist since he would anyway expect quantum theory
to be consistent with instrumentalist argumentation, it does require
some explanation.  And, third, we obtain that the most general evolution
is that of a
superoperator.  This is capable of taking pure states to mixed states.
Hence, collapse interpretations of quantum theory could be incorporated
into this structure.

\vspace{6mm}

\noindent{\Large\bf Acknowledgements}

\vspace{6mm}

I am very grateful to Chris Fuchs for discussions that motivated this
work and to Jeremy Butterfield, Philip Pearle, Terry Rudolph, and Jos
Uffink for comments. This work is funded by a Royal Society University
Research Fellowship.

\vspace{6mm}

\noindent{\Large\bf Appendix 1}

\vspace{6mm}

We will prove that the property
\begin{equation}\label{appconv}
f(\lambda {\bf p}_A + (1-\lambda) {\bf p}_B)=
\lambda f({\bf p}_A) + (1-\lambda) f({\bf p}_B),
\end{equation}
where $\lambda\geq 0$, implies that
\begin{equation}\label{linearproperty}
f({\bf p})=\sum_\alpha a_\alpha f({\bf p}_\alpha)
\end{equation}
where
\begin{equation}\label{palphastate}
{\bf p}=\sum_\alpha a_\alpha {\bf p}_\alpha
\end{equation}
if
\[ {\bf p}_\alpha, {\bf p}\in S  ~~~{\rm for~all}~~~ \alpha\]
for all $a_\alpha$ where $S$ is the set of allowed ${\bf p}$.
First note that putting ${\bf p}_A={\bf 0}$ gives
\begin{equation}
f(\lambda {\bf p})=\lambda f({\bf p})
\end{equation}
for $ 0\leq \lambda \leq 1$.  We can write $\gamma=1/\lambda$ and ${\bf
p}'' = {\bf p}/\lambda$.  Then we obtain
\begin{equation}
f(\gamma {\bf p}'')=\gamma f({\bf p}'')
\end{equation}
where $1\leq\gamma$.  Hence,
\begin{equation} \label{prop}
f(\nu {\bf p})=\nu f({\bf p})
\end{equation}
if $ \nu \geq 0$.  This only follows from (\ref{appconv}) if ${\bf p}, \nu{\bf
p} \in S$.  However, if this is not the case, then the equation does not
correspond to any physical situation. Hence, we are free to impose that
(\ref{prop}) is true for all ${\bf p}$.  In those cases where
${\bf p}, \nu{\bf p} \in S$ is not satisfied the equation has no
physical significance anyway.

Let $f_I$ pertain to that measurement that simply checks to see that a
non-null result has been recorded (we call this the identity
measurement). We will
write $f_I({\bf p})=\mu$.  We define the normalized state $\tilde{\bf
p}$ by $\mu\tilde{\bf p}={\bf p}$ such that $f_I(\tilde{\bf p})=1$
(using (\ref{prop})).

We can normalize each of the states in (\ref{palphastate}) such that
\begin{equation}\label{palphastatenorm}
\mu\tilde{\bf p}=\sum_\alpha a_\alpha\mu_\alpha \tilde{\bf p}_\alpha
\end{equation}
We are free to choose the fiducial measurement corresponding to the first
component of the state vector ${\bf p}$ to be the identity measurement.
Hence, reading off the first component from (\ref{palphastatenorm}) we
obtain
\begin{equation}
\mu= \sum_\alpha a_\alpha\mu_\alpha
\end{equation}
Let $\alpha\in A_{\pm}$ if $a_\alpha$ is $\pm$ve and define
\begin{equation}
\nu=\mu+\sum_{\alpha\in A_-} |a_\alpha|\mu_\alpha
=\sum_{\alpha\in A_+} a_\alpha \mu_\alpha
\end{equation}
We can rearrange (\ref{palphastatenorm})
\begin{equation}
{\mu\over\nu} \tilde{\bf p} + \sum_{\alpha\in A_-} {|a_\alpha
|\mu_\alpha \over \nu}\tilde{\bf p}_\alpha
=\sum_{\alpha\in A_+} {a_\alpha \mu_\alpha \over \nu}\tilde{\bf p}_\alpha
\end{equation}
Each coefficient is positive and the coefficients on each side add to
1. Hence we can apply (\ref{appconv})
\begin{equation}
{\mu\over\nu} f(\tilde{\bf p}) + \sum_{\alpha\in A_-} {|a_\alpha
|\mu_\alpha \over \nu}f(\tilde{\bf p}_\alpha)
=\sum_{\alpha\in A_+} {a_\alpha \mu_\alpha \over \nu}f(\tilde{\bf
p}_\alpha)
\end{equation}
Rearranging this using (\ref{prop}) gives (\ref{linearproperty}) as required.

We see that (\ref{linearproperty}) holds whenever the
arguments of $f$ in each term correspond to physical states.  If these
arguments do not all correspond to physical states then the equation does
not correspond to any physical situation. For mathematical simplicity we
will impose that (\ref{linearproperty}) still holds in such cases.

\vspace{6mm}

\noindent{\Large\bf Appendix 2}

\vspace{6mm}

In this appendix we show that any strictly increasing function having
the completely multiplicative property
\begin{equation}\label{commult}
K(mn)=K(m)K(n),
\end{equation}
where $n$ takes only positive integer values,
is of the form $K(n)=n^\alpha$.  First put $m=n=1$ into
(\ref{commult}). We obtain that $K(1)=0,1$. Put $m=1$ into
(\ref{commult}). If $K(1)=0$ then $K(n)=0$ for all $n$. But this is not
strictly increasing.  Hence we must have $K(1)=1$. The argument $n$ can
be factorized into primes: $n=p_1^{k_1}p_2^{k_2}\dots$ where $p_i$ is
the $i$th prime and the $k_i$'s are integers. It follows from the
completely multiplicative property that
\begin{equation}
K(n)=\prod_i K^{k_i}(p_i)
\end{equation}
Hence, the function $K(n)$ is completely determined by its values at the
primes.  Now consider two primes $p$ and $q$.  Define $\alpha$ by
\begin{equation}
K(p)=p^\alpha
\end{equation}
Note that $K(p)>1$ since $K(n)$ is a strictly increasing function and
hence $\alpha>0$.
Define $a$ by
\begin{equation}
K(q)=aq^\alpha
\end{equation}
Introduce the integer $t$ which we will allow to take any positive
value. Then define $s$ by
\begin{equation}\label{order}
p^s> q^t > p^{s-1}
\end{equation}
From the fact that $K(n)$ is strictly increasing we have
\begin{equation}
K(p^s)> K(q^t) > K(p^{s-1})
\end{equation}
Hence,
\begin{equation}\label{strictorder}
p^{\alpha s} > a^t q^{\alpha t} > p^{\alpha (s-1)}
\end{equation}
Define $\widetilde{s}$ by
\begin{equation}\label{codge}
p^{\widetilde{s}} = q^t
\end{equation}
Comparing with (\ref{order}) we have
\begin{equation}
\widetilde{s}+1 > s>\widetilde{s}> s-1>\widetilde{s}-1
\end{equation}
Hence, (\ref{strictorder}) gives
\begin{equation}
p^{\alpha (\widetilde{s}+1)} > a^t q^{\alpha t} >
p^{\alpha (\widetilde{s}-1)}
\end{equation}
(we have used the fact that $\alpha>0$).
Using (\ref{codge}) we obtain
\begin{equation}
p^\alpha > a^t > p^{-\alpha}
\end{equation}
This must be true for all $t$.  However, $a^t$ can only be bounded from
above and below if $a=1$.  Hence, $K(q)=q^\alpha$.  This applies to any
pair of primes, $p$ and $q$, and hence $K(n)=n^\alpha$.

\vspace{6mm}

\noindent{\Large\bf Appendix 3}

\vspace{6mm}

In this appendix we will prove a number of related important results
some of which are used in the main part of the paper.

The set of reversible transformations is represented by the set,
$\Gamma^{\rm reversible}$, of invertible matrices $Z$ in $\Gamma$ whose
inverses are also in $\Gamma$.  These
clearly form a representation of a group.  In fact, since, by Axiom 5,
this group is
continuous and the vectors ${\bf p}$ generated by the action of
the group remain bounded, $\Gamma^{\rm reversible}$ is a representation
of a compact Lie group.   It can be sown that all real representations
of a compact Lie group are equivalent (under a basis change) to a real
orthogonal representation \cite{boerner}.  Let us perform such a basis
change. Under this basis change assume that $Z\in\Gamma$ is transformed
to $Y\in \Omega$ and ${\bf p}_S \in S$ is transformed to ${\bf q}\in Q$.
The formula $p_{\rm meas}={\bf r}\cdot {\bf p}$ becomes
\begin{equation}
p_{\rm meas}= {\bf s}\cdot {\bf q}
\end{equation}
where ${\bf s}$ now represents the measurement (and is obtainable from
${\bf r}$ by a basis change).  If a transformation device is present
then we have
\begin{equation}
p_{\rm meas}= {\bf s}^T  Y {\bf q}
\end{equation}
We can regard $Y$ as transforming the state or, alternatively, we can
regard it as part of the measurement apparatus.  In this case we have
${\bf s}\rightarrow Y^T{\bf s}$.  If we now restrict our attention to
reversible transformations then $Y\in \Omega^{\rm reversible}$.  But
this is an orthogonal representation and hence $Y^T \in
\Omega^{\rm reversible}$.  Therefore, with this representation, both
states ${\bf q}$ and measurements ${\bf s}$ are acted on by elements of
$\Omega^{\rm reversible}$.

\vspace{4mm}

\noindent{\bf A3.1}

\vspace{4mm}

In this Appendix section we will prove that
\begin{equation}
Z_W^T{\bf r}^{I_W}={\bf r}^{I_W} ~~~{\rm for~all} ~~~Z_W\in\Gamma_W^{\rm
reversible}
\end{equation}
where ${\bf r}^{I_W}$ is the identity measurement for the subspace $W$
and $\Gamma_W^{\rm reversible}$ is the set of reversible
transformations which map states in the subspace $W$ to states in $W$ --
such transformations must exist by Axiom 3.
We can work in the basis for which the transformations are orthogonal
introduced above.  Then we wish to prove
\begin{equation}\label{sinvar}
Y^T_W{\bf s}^{I_W}={\bf s}^{I_W} ~~~{\rm for~all}~~~Y_W\in\Omega_W^{\rm
reversible}
\end{equation}
Working in this basis we can write any state in the subspace $W$ as
\begin{equation}\label{qax}
{\bf q}= a {\bf s}^{I_W} + {\bf x}
\end{equation}
where ${\bf x}$ is orthogonal to ${\bf s}^I$.
The normalization of this state is fixed by $a$. Let $K_W$ be the number
of degrees of freedom associated with the subspace $W$.
 Once the normalization
coefficient has been fixed there are $K_W-1$ degrees of freedom left
corresponding to the $K_W-1$ dimensions of the vector space orthogonal to
${\bf s}^{I_W}$  for states in $W$ which is spanned by possible ${\bf
x}$.  There must be
at least one direction in this vector space for which both ${\bf x}$
and $\gamma{\bf x}$, where $\gamma\not=1$,
are permissible vectors (corresponding to allowed states).  To see this
assume the contrary. Thus assume that for each direction ${\bf x}/|{\bf
x}|$ there is only one allowed length of vector. Such a constraint would
remove one degree of freedom leaving $K_W-2$ degrees of freedom which
contradicts our starting point that there are
$K_W-1$ degrees of freedom
associated with states with a particular normalization coefficient.
Consider such an ${\bf x}$ for which $\gamma{\bf x}$ is also
permissible. Now
\begin{equation}
{\bf s}^{I_W}\cdot Y_W {\bf q}={\bf s}^{I_W}\cdot{\bf q}
\end{equation}
since the reversible transformation $Y_W$ does not change the
normalization coefficient of the state and ${\bf q}$ is in $W$ both
before and after the transformation. Using (\ref{qax}) this becomes
\begin{equation}
a{\bf s}^{I_W}\cdot Y_W {\bf s}^{I_W} + {\bf s}^{I_W}\cdot Y_W {\bf x}
 =a{\bf s}^{I_W}\cdot{\bf s}^{I_W}
\end{equation}
This equation must also apply when ${\bf x}$ is replaced by $\gamma{\bf
x}$.
\begin{equation}
a{\bf s}^{I_W}\cdot Y_W {\bf s}^{I_W} + \gamma{\bf s}^{I_W}\cdot Y_W {\bf x}
 =a{\bf s}^{I_W}\cdot{\bf s}^{I_W}
\end{equation}
Subtracting these two equations tells us that the second term on the LHS
vanishes. Hence
\begin{equation}
{\bf s}^{I_W}\cdot Y_W {\bf s}^{I_W}
 ={\bf s}^{I_W}\cdot{\bf s}^{I_W}
\end{equation}
Now, the transformation $Y_W$ is orthogonal and hence length preserving
and thus (\ref{sinvar}) follows.

It follows that
\begin{equation}
Y^T{\bf s}^I ={\bf s}^I ~~~ {\rm for~all}~~~ Y\in\Omega^{reversible}
\end{equation}
where ${\bf s}^I$ is the identity measurement in this new basis (written
as ${\bf r}^I$ in the usual basis). This property is to be expected
since reversible transformations leave the normalization coefficient of
a state unchanged.

\vspace{4mm}

\noindent{\bf A3.2}

\vspace{4mm}

It is clearly the case that
\begin{equation}
Z_W {\bf p}\in W ~~~ {\rm if} ~~~ {\bf p}\in W  ~~~
{\rm and}~~~   Z_W\in\Gamma_W^{\rm reversible}
\end{equation}
It is also the case that
\begin{equation}
Z_W {\bf p}\in\overline{W} ~~~{\rm if}~~~ {\bf p}\in \overline{W}  ~~~
{\rm and}~~~   Z_W\in\Gamma_W^{\rm reversible}
\end{equation}
where $\overline{W}$ is the complement subspace of  $W$.
This follows immediately since ${\bf p}\in\overline{W}$ iff ${\bf
r}^{I_W}\cdot{\bf p}=0$.  But if this is true then, since $Z_W^T{\bf
r}^{I_W}={\bf r}^{I_W}$, it is also true that ${\bf
r}^{I_W}\cdot Z_W{\bf p}=0$. Hence, $Z_W{\bf p}\in\overline{W}$.

\vspace{4mm}

\noindent{\bf A3.3}

\vspace{4mm}

We will now prove that we can choose $Z_W$ such that $Z_W{\bf p}_n={\bf
p}_n$ for $n\in\overline{W}$.  Define $W'(m)$ to be the set containing
all the elements of $W$ plus the first $m$ elements of $\overline{W}$.
Consider only states constrained to the subspace $W'(1)$ and consider
the set $\Gamma^{\rm reversible}_{W'(1)}$ of reversible transformations
which map states in $W'(1)$ back into $W'(1)$.
The subspace $W$ is a subspace of $W'(1)$. Hence, by Axiom 3, there must
exist a subset of $\Gamma^{\rm reversible}_{W'(1)}$ which map states in
$W$ back into $W$.  By the result in A3.2 these transformations must
leave the basis state ${\bf p}_{m_1}$ unchanged (where $m_1$ is the
first entry of $\overline{W}$) since this is the only normalized state
in $W'(1)$ and the complement of $W$.  We can now run the same argument
taking $W'(2)$ to be our system and so on. In this way we establish that
we can find a transformations $Z_W$ which have the desired property.

\vspace{4mm}

\noindent{\bf A3.4}

\vspace{4mm}

In this appendix subsection we show that one possible choice of fiducial
measurements are those identified in Section \ref{choosing}.
Consider the set $\Gamma_{mn}^{\rm reversible}$ of reversible
transformations that transform states in the subspace $W_{mn}$ to states
in the same subspace (where $W_{mn}$ is the subspace associated with the
$m$th and the $n$th basis vectors).
It follows from the property established in A3.2 that
\begin{equation}\label{compred}
{\bf r}^T_n Z_{mn} {\bf p} = 0 ~~~~{\rm if}~~~~ {\bf p}\in
\overline{W}_{mn}
\end{equation}
We can regard the transformation device as part of the measurement
apparatus (rather than regarding it as acting on the state).  In this
case we have
\begin{equation}
{\bf r}_n\rightarrow Z^T_{mn} {\bf r}_n
\end{equation}
We can choose two particular transformations $Z_{mnx}$ and $Z_{mny}$ to
provide us with the two extra needed fiducial measurements, ${\bf
r}_{mnx}$ and ${\bf r}_{mny}$ respectively,  for each two-dimensional
subspace.  The vectors ${\bf r}_m$, ${\bf r}_n$, ${\bf r}_{mnx}$, and
${\bf r}_{mny}$ must be linearly independent.  It follows from the
fact that, for this subspace, the group of transformations is
equivalent, under a basis change, to the full group of orthogonal rotations
in three dimensions that we
can choose $Z_{mnx}$ and $Z_{mny}$ such that this is the case.
From (\ref{compred}) we have
\begin{equation}\label{useful}
{\bf r}_{mnx} \cdot {\bf p} = 0 ~~~{\rm if}~~~ {\bf p}\in \overline{W}_{mn}
\end{equation}
and similarly for ${\bf r}_{mny}$.

We will now prove that the $N^2$ vectors chosen in this way are linearly
independent. We can do this by showing that each
measurement yields information about the state that none of the others do.
First, the vectors ${\bf r}_n$ are linearly independent of
each other since there exists a vector (namely ${\bf p}_m$) having
non-zero overlap with any given ${\bf r}_m$ which has zero overlap with all the
other ${\bf r}_n$. Now we add two fiducial vectors, ${\bf r}_{mnx}$ and
${\bf r}_{mny}$, to each two-dimensional
subspace $W_{mn}$ that are, by construction, linearly independent of the
basis vectors already in that subspace.  Since the fiducial measurements
pertaining to one
such two-dimensional subspace yield no information about states in any
other non-overlapping two-dimensional subspace (because of
(\ref{useful}))they must be linearly independent of the fiducial
measurements in those
non-overlapping subspaces.  What about overlapping two-dimensional
subspaces? Consider performing the measurement
${\bf r}_{mnx}$ on ${\bf p}$ in $W_{mn'}$ where $n'\not=n$.  Since
${\bf p}$ is in $W_{mn'}$ it follows from Axiom 3 that the measurement
${\bf r}_{mnx}$ must be equivalent to some measurement in this subspace
(though only for states in this subspace).  Now, if the state is
actually the basis state ${\bf p}_{n'}$ then zero probability would be
recorded.  This means that the measurement ${\bf r}_{mnx}$, when
regarded as a measurement on $W_{mn'}$ is actually equivalent to a
measurement just on the one-dimensional subspace $W_m$.
Hence, ${\bf r}_{mnx}$ does not yield any information about states in the
subspace $W_{mn'}$ that is not given by ${\bf r}_m$ and therefore the
measurements ${\bf r}_{mn'x}$ and ${\bf r}_{mn'y}$ are linearly
independent of it.  Hence, the $N^2$ fiducial measurements are all
necessary to determine the state and are therefore linearly independent.

\vspace{4mm}

\noindent{\bf A3.5}

\vspace{4mm}

In this appendix subsection we show that the map between a pure state
and that pure measurement identifying it is linear and invertible
(recall that pure measurements are defined to be those measurements
which can be obtained by acting on the basis measurement ${\bf r}_1$
with a reversible transformation).
Using the basis for which reversible transformations are orthogonal (see
introduction to this appendix) we can put
\begin{equation}
{\bf q}= a {\bf s}^I + {\bf u}
\end{equation}
for the state, and
\begin{equation}
{\bf s}= b {\bf s}^I + {\bf v}
\end{equation}
for the measurement
where ${\bf u}$ and ${\bf v}$ are orthogonal to ${\bf s}^I$.  Since
$Y^T {\bf s}^I={\bf s}^I$ and since the group of reversible
transformations, $\Omega^{\rm reversible}$, is orthogonal it follows
that $Y {\bf s}^I={\bf s}^I$.   Hence, transformations only effect the
components of ${\bf q}$ and ${\bf s}$ orthogonal to ${\bf s}^I$.
Using $p_{\rm meas}={\bf r}\cdot {\bf p}={\bf s}\cdot{\bf q}$ we obtain
\begin{equation}
p_{\rm meas}=k + {\bf v}\cdot{\bf u}
\end{equation}
where $k=ab{\bf s}^I\cdot{\bf s}^I$.

Now assume that the
pure measurement represented by ${\bf s}$ identifies the pure state
represented by ${\bf q}$. Then $k+{\bf u}\cdot {\bf v}=1$.  This
probability cannot be increased by any transformation device. Hence,
\begin{equation}\label{vYu}
{\bf v}^T Y {\bf u} \leq {\bf v}^T {\bf u} ~~~{\rm for~all}~~~
Y\in\Omega^{\rm reversible}
\end{equation}
Since the orthogonal transformation $Y$ is length preserving it would
appear that the only way to satisfy this condition is
if ${\bf v}$ is parallel to ${\bf u}$.  This is indeed the case and is
proven at the end of this appendix subsection.
Hence, we can say that the state ${\bf q}= a {\bf s}^I + {\bf
u}$ is identified by the measurement ${\bf s}= b {\bf s}^I + c{\bf u}$.
Now apply this result to the basis state ${\bf q}_1$ (this corresponds
to ${\bf p}_{1}$) and the basis measurement ${\bf s}_1$ (this
corresponds to ${\bf r}_{1}$).  Let $C$ be the linear map that performs
scalar multiplication by a factor $\mu$ in the ${\bf s}^I$ direction and
by a factor $\nu$ in the subspace orthogonal to ${\bf s}^I$.  We can
apply $C$ to ${\bf q}$ and $C^{-1}$ to ${\bf s}$ such that
${\bf s}_1={\bf q}_1=\alpha{\bf s}^I + \beta{\bf u}_1$ by appropriate choice of
the factors $\mu$ and $\nu$.  The maps $C$ and $C^{-1}$ commute with the
orthogonal transformations $\Omega^{\rm reversible}$. Hence, in general,
the pure state ${\bf q}=\alpha{\bf s}^I + Y\beta{\bf u}_1$ is identified
by the pure measurement ${\bf s}=\alpha{\bf s}^I + Y\beta{\bf u}_1$
(i.e. represented by the same vector) as $YY^T=I$.
Since the basis change and the
maps $C$ and $C^{-1}$ are all linear and invertible it follows that the map from
pure states to the pure measurements identifying them is linear and
invertible.

As promised, we will now prove that ${\bf v}$ is parallel to ${\bf u}$
when a pure
measurement ${\bf s}$ identifies a pure state ${\bf q}$.  First consider
the basis measurement ${\bf s}_1$ and the basis state ${\bf q}_1$ it
identifies.  Let $V_{mn}$ be the vector space spanned by the fiducial
measurement
vectors ${\bf s}_m$, ${\bf s}_n$, ${\bf s}_{mnx}$ and ${\bf s}_{mny}$
associated with the $mn$ subspace.  It follows from $A3.4$ that these
vector spaces span the full $N^2$ dimensional vector space.  The state
${\bf q}_1$ can have no projection into the vector space $V_{mn}$ if
$m,n\not=1$ (since ${\bf s}\cdot{\bf q}_1=0$ for ${\bf s}$ associated
with the $mn$ subspace).  Let $\widetilde{V}$ be the vector space
spanned by the vector spaces $V_{1n}$ for $n=1$ to $N$.  It follows from
the fact that ${\bf q}_1$ has no projection into $V_{mn}$ for
$m,n\not=1$ that ${\bf q}_1$
is in $\widetilde{V}$. Now the vector ${\bf s}^I=\sum_n {\bf s}_n$ is
clearly in $\widetilde{V}$. Let $\widetilde{V}'$ be the vector space in
$\widetilde{V}$ orthogonal to ${\bf s}^I$.  We can write ${\bf q}_1=a{\bf
s}^I+{\bf u}_1$ where ${\bf u}_1$ is in the vector space
$\widetilde{V}'$.  Similarly, we can write ${\bf s}_1=b{\bf s}^I+{\bf
v}_1$.
Define $V'_{1n}$ as the vector space spanned by the fiducial measurement
vectors ${\bf v}_1$, ${\bf v}_n$, ${\bf v}_{1nx}$, and ${\bf v}_{1ny}$
associated with the subspace $1n$.  The vector spaces $V'_{1n}$ for
$n=1$ to $N$ span $\widetilde{V}'$.  Consider
orthogonal transformations $Y_{1n}$
which leave states in the $1n$ subspace. They will also transform
measurements pertaining to the $1n$ subspace to measurements still
pertaining to this subspace (and thus still in $V'_{1n}$).  Since ${\bf
v}_1$ and $Y_{1n}^T{\bf v}_1$ are both in $V'_{1n}$ we can write
(\ref{vYu}) as
\begin{equation}\label{vYu1n}
{\bf v}_1^T Y_{1n} {\bf u}_1^{1n} \leq {\bf v}_1^T {\bf u}_1^{1n}
\end{equation}
where ${\bf u}_1^{1n}$ is the component of ${\bf u}_1$ in $V'_{1n}$.
The vector space $V'_{1n}$ is three dimensional
and the action of the group of orthogonal
transformations in the $1n$ subspace on ${\bf u}_1$ is to sweep out a
sphere (since these transformations are length preserving).
Hence, condition (\ref{vYu1n}) can only be satisfied for all rotations
$Y_{1n}$ if ${\bf u}_1^{1n}$ is parallel to ${\bf v}_1$.
The vector spaces $V'_{1n}$ span all of $\widetilde{V}'$ and
hence ${\bf u}_1$ has no component which is perpendicular to ${\bf
v}_1$.  This means that ${\bf v}_1$ is parallel to ${\bf u}_1$.  We
complete the proof by noting that a
general pure measurement can be written ${\bf v}=Y{\bf v}_1$ and
identifies the pure state ${\bf u}=Y{\bf u}_1$.

\vspace{4mm}

\noindent{\bf A3.6}

\vspace{4mm}

It is easy to prove that $D=D^T$.  We chose a set of pure fiducial
states and we chose the fiducial measurements to be the set of pure
measurements that identify these states.  Hence, if we represent the
fiducial states by a set of vectors ${\bf q}^l$ then, as proven in A3.5,
we can represent the fiducial measurements by the the same vectors ${\bf
s}^k={\bf q}^k$.  The matrix element $D_{kl}$ is equal to the probability when
the $k$th fiducial measurement is performed on the $l$th fiducial state.
This is equal to ${\bf q}^k\cdot{\bf q}^l$ and hence $D=D^T$.

\vspace{4mm}

\noindent{\bf A3.7}

\vspace{4mm}

Now we will show that the basis measurements ${\bf r}_{n}$ are all
pure and, therefore, that all the fiducial measurements of $A3.4$ are
pure.  Consider first the case where $N=2$. Then $K=4$.  The normalized
states (and hence pure states) live in a three dimensional space (since
we can eliminate one variable by normalization).  Hence, orthogonal
transformations can be regarded as rotations about an axis.
We can write the basis
states as
\begin{equation}
{\bf q}_1=\alpha {\bf s}^I + \beta{\bf u}_1
\end{equation}
\begin{equation}
{\bf q}_2=\alpha {\bf s}^I - \beta{\bf u}_1
\end{equation}
This follows since there exists a continuous orthogonal
transformation which takes ${\bf q}_1$ to ${\bf q}_2$.  This can be
regarded as a rotation around a great circle.  The orthogonal state ${\bf
q}_2$ must correspond to the opposite point on this circle where ${\bf
u}=-{\bf u}_1$ since this is the point at which ${\bf s}_1\cdot {\bf q}$
stops decreasing and starts increasing again.
Now, we have already that
\begin{equation}
{\bf s}_1=\alpha {\bf s}^I + \beta{\bf u}_1
\end{equation}
We have not yet proven that ${\bf s}_2$ (corresponding to ${\bf
r}_{2}$) is pure. However, we know that ${\bf s}_1+{\bf s}_2={\bf s}^I$
so we can write
\begin{equation}
{\bf s}_2=\alpha' {\bf s}^I - \beta{\bf u}_1
\end{equation}
with $\alpha+\alpha'=1$.  It then follows from ${\bf s}_1\cdot{\bf q}_2=
{\bf s}_2\cdot{\bf q}_1=0$ that $\alpha=\alpha'=1/2$.  Hence, ${\bf
s}_2$ is pure.   This proof can be applied to the general $N$ case by
considering only a two dimensional subspace.  It follows from Axiom 3
that there must exist a set of invertible transformations which
transform states in the $1n$ subspace to states in the same subspace.
As shown in A3.1, these leave ${\bf s}^{I_{W_{1n}}}$ invariant (this is
the identity
measurement vector for the $1n$ subspace). Hence, we can replace ${\bf
s}^I$ by ${\bf s}^{I_{W_{1n}}}$ throughout the above proof if we are only
considering transformations in this subspace.  It follows that we can
transform ${\bf s}_1$ to ${\bf s}_n$ and hence the basis measurements
are all pure.

Hence we can transform ${\bf s}_1$ to any ${\bf s}_{mnx}$ by first
transforming by a reversible transformation to ${\bf s}_n$ and then
applying the reversible transformation of A3.4 to obtain ${\bf
s}_{mnx}$. Similar remarks apply to ${\bf s}_{mny}$.

\vspace{6mm}

\noindent{\Large\bf Appendix 4}

\vspace{6mm}

In this appendix we will show that the projector
$\hat{P}^A_i\otimes\hat{P}^B_j$ can correspond to the measurement of the joint
probability of obtaining a positive outcome for fiducial measurements
$i$ at $A$ and $j$ at $B$, and to the state when the $i$th fiducial
state is prepared at $A$ and the $j$th fiducial state is prepared at
$B$. First, note that we can prepare $N_AN_B$ distinguishable states
for the composite system by preparing basis state $m$ at $A$ and basis
state $n$ at $B$.  Since the composite system has $N=N_AN_B$ this
represents a complete set of basis states.  Further, since $K(1)=1$ all
basis states must be pure (as noted at the end of Section \ref{secKNr}).
Hence, we can choose these basis states to correspond to the basis states of
our Hilbert space $|mn\rangle$ or, equivalently,
$|m\rangle\otimes|n\rangle$.  As operators these basis states are
$\hat{P}^A_m\otimes\hat{P}^B_n$ where $m$ ($n$) only runs over the first
$N_A$ ($N_B$) values (the remaining values corresponding to the other
fiducial projectors).

Now consider the $N_B$ dimensional subspace
with basis states $|1\rangle\otimes|n\rangle$ ($n=1$ to $N_B$). This
subspace corresponds to the case where system $A$ is prepared in basis
state $1$ and system $B$ is prepared in any state.  A full
set of fiducial projectors can be formed for this subspace.  These will
take the form $\hat{P}^A_1\otimes\hat{P}^B_l$ where $l=1$ to $K_B$ (i.e.
runs over the all values, not just the basis labels).  We can do the
same for the case where basis state 2 is prepared at $A$.  Then we have
the fiducial projectors $\hat{P}^A_2\otimes\hat{P}^B_l$ for the subspace
$2n$ ($n=1$ to $N_B$). Indeed, we can do this for the general case in
which the basis state $m$ is prepared at $A$.  Now consider the pure
state $\hat{P}^A_1\otimes\hat{Q}^B$ where $\hat{Q}^B$ is some arbitrary
projector at $B$.  This state is in the $1n$ ($n=1$ to $N_B$) subspace
and we can perform the fiducial measurements
$\hat{P}^A_1\otimes\hat{P}_l^B$ in this subspace to fully characterize
this state.  The probabilities obtained in making these fiducial
measurements will be the same as if we prepared the state
$\hat{P}^A_2\otimes\hat{Q}^B$ and made the fiducial measurements
$\hat{P}^A_2\otimes\hat{P}_l^B$ and hence this corresponds to the same
preparation at $B$.  Hence, in general the projector
$\hat{P}^A_m\otimes\hat{Q}^B$ corresponds to preparing the basis state
$\hat{P}_m$ at $A$ and the general pure state $\hat{Q}^B$ at $B$.
Now consider the subspace spanned by
the projectors $\hat{P}^A_m\otimes\hat{Q}^B$ ($m=1$ to $N_A$) in which
we prepare $\hat{Q}^B$ at $B$.  A fiducial set for this subspace is
$\hat{P}^A_k\otimes\hat{Q}^B$ where $k=1$ to $K_A$.  If these fiducial
measurements are made on a state $\hat{R}^A\otimes\hat{Q}^B$ where
$\hat{R}^A$ is a projector at $A$ then we would get the same results as
if the fiducial measurements $\hat{P}^A_k\otimes\hat{Q'}^B$ were made on
the state $\hat{R}^A\otimes\hat{Q'}^B$.  Hence, in both cases the
preparation at $A$ is the same.  Thus, the pure state
$\hat{R}^A\otimes\hat{Q}^B$ corresponds to the case where a particular
pure state $\hat{R}^A$ is is prepared at $A$ and the pure state
$\hat{Q}^B$ is prepared at $B$.  An analogous argument to that above can
be used to show that, regarded as a measurement, the
projector $\hat{R}^A\otimes\hat{Q}^B$ corresponds to
measuring the joint probability with setting $\hat{R}^A$ at end $A$ and
setting $\hat{Q}^B$ at end $B$.   Applied to the fiducial projectors,
$\hat{P}_k^A\otimes\hat{P}_l^B$, this proves our result.

\end{document}